\documentclass[
  aps,
  prb,
  twocolumn,
  superscriptaddress,
]{revtex4-2}

\usepackage{graphicx}
\usepackage{dcolumn}
\usepackage{bm}
\usepackage{epstopdf}
 \usepackage[usenames,dvipsnames]{pstricks}
 \usepackage{epsfig}
 \usepackage{pst-grad} 
 \usepackage{pst-plot} 
 \usepackage[space]{grffile} 
 \usepackage{etoolbox} 
 \makeatletter 
 \patchcmd\Gread@eps{\@inputcheck#1 }{\@inputcheck"#1"\relax}{}{}
 \makeatother
\usepackage{booktabs}
\usepackage{boxhandler}
\usepackage{amsmath, amssymb}
\usepackage{bbold}
\usepackage{overpic}
\usepackage{natbib}
\usepackage{bm}
\usepackage[whole]{bxcjkjatype}
\usepackage{mathtools}

\usepackage{floatrow}
\usepackage[utf8]{inputenc}
\usepackage{braket}
\usepackage[cmintegrals]{newtxmath}
\usepackage{multirow}
\usepackage{enumitem}
\usepackage{pifont}
\usepackage{url}
\usepackage{algorithm}
\usepackage{algpseudocode}
\usepackage{subcaption}
\usepackage{hyperref}

\hypersetup{
  colorlinks   = true,    
  urlcolor     = blue,    
  linkcolor    = blue,    
  citecolor    = blue      
}
\usepackage[sort]{cleveref}
\algnewcommand{\FUNCTION}[2]{%
  \algorithmicfunction\ \textproc{#1}(#2)%
}
\algnewcommand{\ENDFUNCTION}{%
  \algorithmicend\ \algorithmicfunction%
}
\algdef{SE}[FUNCTION]{Function}{EndFunction}%
[2]{\FUNCTION{#1}{#2}}{\ENDFUNCTION}%

\renewcommand{\vec}[1]{\bm{#1}}
\DeclareMathOperator{\Tr}{Tr}
\renewcommand{\H}{\mathcal{H}}

\begin{document}

\title{Noise mitigation of quantum observables via learning from Hamiltonian symmetry decays}
\author{Javier {Oliva del Moral}}
\email{jolivam@unav.es}
\affiliation{Department of Basic Sciences, Tecnun - University of Navarra, 20018 San Sebastian, Spain}
\affiliation{Donostia International Physics Center, 20018 San Sebastián, Spain.}
\author{Olatz {Sanz Larrarte}}
\affiliation{Department of Basic Sciences, Tecnun - University of Navarra, 20018 San Sebastian, Spain}
\author{Joana Fraxanet}
\affiliation{IBM Quantum, IBM Thomas J Watson Research Center, Yorktown Heights, NY 10598, USA}
\author{Dmytro Mishagli}
\affiliation{IBM Quantum, IBM Research Europe -- Dublin}
\affiliation{IBM Quantum, IBM Research Europe~-- UK}
\author{Josu {Etxezarreta Martinez}}
\email{jetxezarreta@unav.es}
\affiliation{Department of Basic Sciences, Tecnun - University of Navarra, 20018 San Sebastian, Spain}

\begin{abstract}
We present a new quantum error mitigation technique (QEM), called GUiding Extrapolations from Symmetry decayS (GUESS), which exploits Hamiltonian symmetries to improve accuracy of noisy quantum computations. This method is explicitly designed for quantum algorithms that estimate expectation values of observables and consists in learning the extrapolation coefficients from a symmetry observable of the system to then estimate the value of a target observable. Furthermore, we propose a Hamiltonian impurity technique to enforce symmetries allowing the mitigation of local observables of interest. We employ the IBM Heron r2 quantum processing unit '\texttt{ibm\_basquecountry}' to simulate the time evolution of average magnetization and nearest-neighbor correlator observables for transverse field Ising and $XZ$ Heisenberg models in 1D with open boundary conditions. We benchmark the accuracy of our method against baseline Zero Noise Extrapolation (ZNE) and tensor network simulations for systems of $100$ qubits. Remarkably, GUESS achieves a relative error around $10\%$ for circuits containing up to $8000$ CZ gates, while showcasing lower variance than ZNE on average across $20$ observables and requiring only twice the number of shots per observable compared to baseline ZNE. Furthermore, we demonstrate that GUESS enables statistical post-selection based on the outcomes of the symmetry observable, which provides critical information about the quality of the target qubits by means of its mean and variance. These results indicate that GUESS is a powerful QEM technique capable of mitigating utility-scale circuit outcomes, delivering high accuracy and reduced variance for large-scale circuits with minimal quantum overhead.
\end{abstract}

\maketitle
\section{Introduction} \label{Sec: Intro}
Quantum computing aims to revolutionize the paradigm of modern computation by providing super-polynomial advantages respect to classical computers in solving a special set of problems ranging from quantum mechanical simulation~\cite{Feynman1982} to purely algebraic problems such as integer factorization~\cite{Shor_1997}. Unfortunately, the existence of quantum noise significantly hinders the performance of quantum computers, introducing errors in the computations that cause the results to be inaccurate~\cite{9201447,Fowler_2012}. In practice, quantum noise imposes a fundamental limit to the size, both in terms of circuit depth and qubit number, of the quantum algorithms that can be implemented in noisy quantum computers~\cite{PhysRevA.51.992}.

Quantum Error Correction (QEC) is considered to be the solution to construct fault-tolerant quantum computers capable of executing algorithms of arbitrary size~\cite{Calderbank_1996,PhysRevA.52.R2493,PhysRevLett.77.793}. The celebrated threshold theorem~\cite{10.1145/258533.258579,Kitaev_1997} states that if errors in quantum hardware are reduced below a finite rate, known as the threshold, quantum computations of arbitrary size can be performed reliably using fault-tolerant gadgets. However, many technical challenges must be resolved to construct such machines, including building hardware with sub-threshold error rates, real-time decoding, magic state distillation or providing qubit overheads that seem daunting at this moment in time~\cite{Fowler_2012,decoders,litinski}. For example, recent optimized approaches show that scientific applications that are classically intractable may require hundreds of thousands of qubits~\cite{Kivlichan_2020}, while industrial applications will require millions of qubits~\cite{Lee_2021}. Although theoretical research is ongoing to find alternative codes with more favorable overheads and recent progress gives reason for optimism with respect to construct the first small error corrected quantum computers~\cite{Breuckmann_2021,BBcodes,googlesurf,tourdegross,MagicExpQuera}, the challenge of realizing fault-tolerant large-scale quantum computing is considerable and remains as a long-term goal.

\begin{figure*}[th]
    \includegraphics[width=\textwidth]{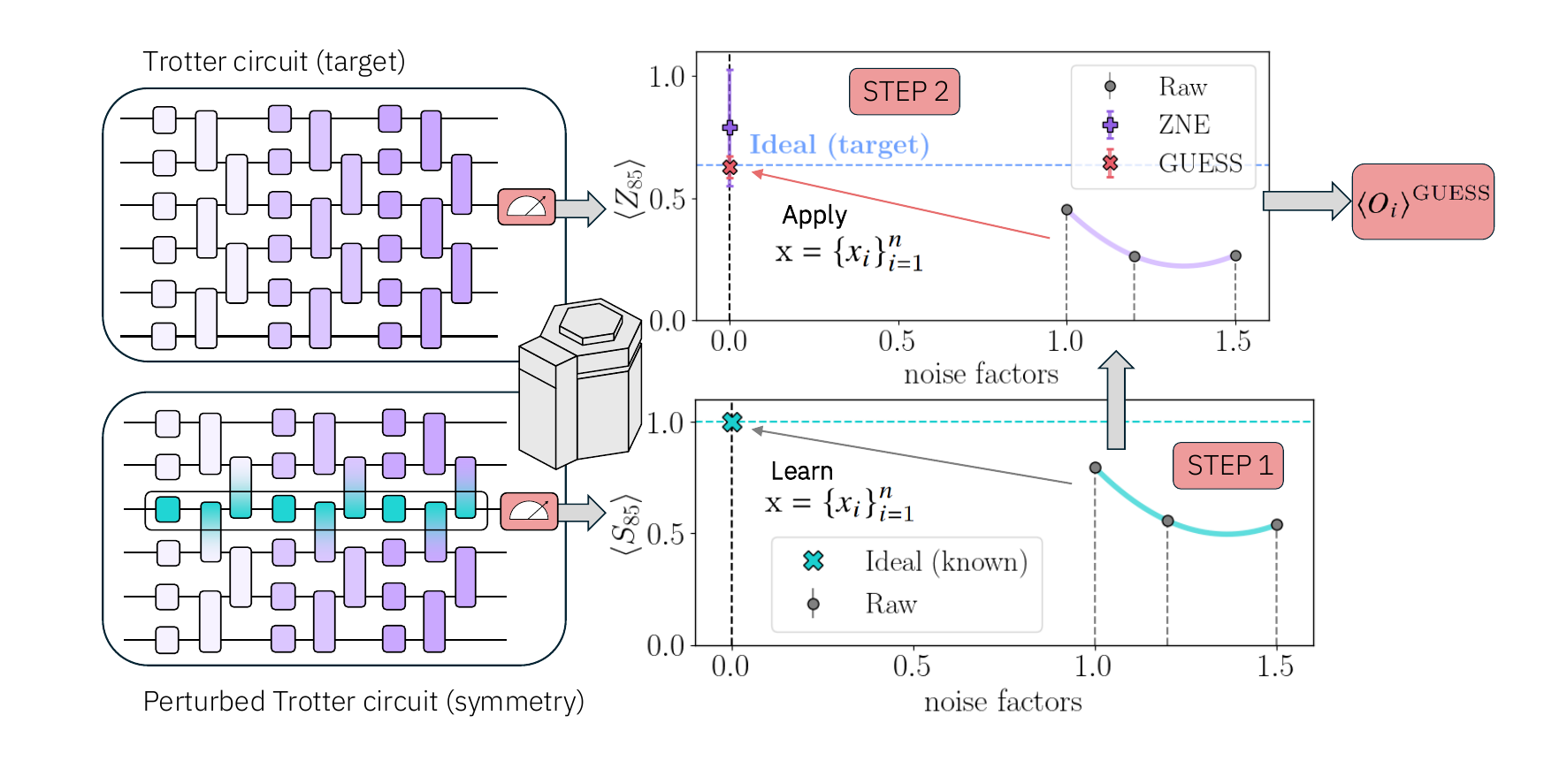}
    \caption{\textbf{Schematic representation of the GUESS method.} The top left panel shows a representative Trotter circuit used to estimate the time evolved expectation value of the target observable $Z_{85}$ at qubit $85$ for different noise factors $(1, 1.2, 1.5)$. The bottom left panel illustrates the same circuit with a local impurity added, which generates a symmetry observable at the same qubit, $S_{85}$. In STEP 1, we measure expectation values of the symmetry observable at each noise factor and we use them together with the known ideal value $\langle S_{85} \rangle_{ideal} = 1$ to learn the GUESS coefficients $\{x_i\}_{i=1}^{n}$. In STEP 2, these coefficients are applied to the measured expectation values of $\langle Z_{85} \rangle$ at the same noise factors to estimate its noise‑free value. The GUESS estimate is compared to the ideal (target) result, which here is obtained with tensor networks, and to a standard ZNE extrapolation. We used the IBM heron r2 '\texttt{ibm\_basquecountry}' for simulating the dynamics of an transverse field Ising model at Trotter step $20$.}
    \label{fig: Schematic}
\end{figure*}

In this context, quantum error mitigation (QEM) emerged as a strategy for performing computations with useful on near-term quantum processors~\cite{Cai_2023,ZNETemme}. The QEM paradigm refers to the set of algorithmic techniques and methods that reduce the noise-induced bias in the evaluation of expectation values of observables of interest (e.g. molecular energy levels or magnetization of ferromagnetic materials) by post-processing the outputs from an ensemble of noisy quantum circuits or filtering them using a post-selection algorithm~\cite{Cai_2023}. Although these techniques were conceived for near-term hardware, the community expects those to co-exist with fault-tolerant quantum machines, especially in the early fault-tolerance setting~\cite{MegaquopMachine,MythsQEM,lanes2025, jeon2026}. Zero Noise Extrapolation (ZNE)~\cite{li_benjamin_zne,ZNETemme,Giurgica_Tiron_2020,majumdar2023best,commentZNE} and Probabilistic Error Cancellation (PEC)~\cite{ZNETemme,PEC1,PEC2,PEC3} are the mostly celebrated QEM techniques at this point and have shown the ability to produce accurate results for quantum algorithms of non-trivial sizes~\cite{UtilityIBM,PEC3,GoogleEchoes}.

Although ZNE techniques have proven to be effective in practical scenarios, they still face challenges that need to be addressed. First of all, it is known that complex non-Clifford circuits may lead to multi-exponential decays~\cite{Cai_2023,UtilityIBM,caimultiexp}. As noted in~\cite{Cai_2023}, a precise determination of the extrapolation curve is essential for improving the accuracy of noiseless expectation value estimations. In this sense, while techniques based on Richardson extrapolation theoretically avoid assuming a specific decay curve, they result to be unstable in practice. This instability often motivates the use of simpler models, such as exponential or linear extrapolation, to achieve more reliable results~\cite{UtilityIBM}. Recent research efforts have focused on developing methods to improve the effectiveness and robustness of ZNE~\cite{QEMtech1,QEMtech2,QEMtech3,QEMtech4}.

In this work, we propose a method of learning decoherence laws as functions of noisy expectation values to improve existing error-mitigation techniques such as ZNE. We achieve this by studying noisy dynamics of observables, $S$, that generate continuous symmetries of a Hamiltonian $\H$, \textit{i.e.} that commute with it, $[\H,S]=0$. We will refer to such observables as to symmetries. Hamiltonian symmetries have previously been considered in the literature to discard measurements that violate these symmetries, as these outcomes are nonphysical or do not belong to the right symmetry sector~\cite{Cai_2023,symmetries,cobos}. In contrast, our approach aims to leverage the use of the conserved values to learn the decay of expectation values of the symmetries to then effectively mitigate the target observables.

We also prove that the decay of an observable depends on its weight, defined by the number of non trivial measured operators. Therefore, we need to map the learned noise of the symmetry to the noise suffered by a target observable of the same weight. However, quantum many-body Hamiltonians usually present a low number of symmetries, most of which are global and, thus, decay very fast~\cite{UtilityIBM,Fauseweh2024,PhysRevX.5.021027}. To address this limitation, we propose a perturbation method that enforces a symmetry of a desired weight while preserving the propagation of the noise. We can effectively use the measured expectation values on the perturbed circuits to learn the optimal extrapolation coefficients by solving a least-squares estimation problem, since the ideal value of the enforced symmetry is known. Moreover, the measured symmetry allows us to perform a post-selection algorithm to choose the best qubits based on their corresponding symmetry, since it gives us information about their decay and standard deviation. 

Our method, which we call GUiding Extrapolations from Symmetry decayS (GUESS), is represented in Fig~\ref{fig: Schematic}.  We demonstrate its effectiveness through extensive experiments on Trotterized circuits for quantum time evolution of both the Ising and $XZ$ Heisenberg chains with $100$ qubits, simulating $44$ and $24$ Trotter steps, respectively. These circuits, containing up to $8000$ two-qubit gates, are executed on an IBM Heron r2 QPU. We benchmark GUESS by averaging error-mitigated outcomes over post-selected sets of the best $20$ observables with varying weights, showcasing its robustness across diverse measurement targets.

The paper is organized as follows. In Section~\ref{Sec: ProblemStatement}, we present the QEM problem following by a description of our method to solve it, called GUiding Extrapolations from Symmetry decayS (GUESS) in Section~\ref{Sec: ProjectedLin}. Next, Section~\ref{Sec: Simulations} focuses on how we map the GUESS method into an actual quantum computer, using an IBM Heron r2 QPU '\texttt{ibm\_basquecountry}', followed by Section~\ref{Sec: ExperimentResults}, where we present and analyze the results. Finally, Section~\ref{Sec: Discussion} concludes the paper with an extended discussion of the proposed method and its implications.

\section{Problem statement} \label{Sec: ProblemStatement}

Many quantum computing tasks involve estimating the expectation value of an Hermitian observable of interest $O$ for a quantum state $\rho$, \textit{i.e.} $\langle O\rangle = \Tr{(O\rho)}$. In the ideal setting (\textit{i.e.} without noise), the evolution of a closed quantum system under a Hamiltonian $\H$ is described by the \textit{von Neumann} equation~\cite{VonNeumann}:
\begin{equation}
    \frac{\partial}{\partial t}\rho(t) = -\frac{i}{\hbar}[\H, \rho(t)], \label{eq:von_neumann}
\end{equation}
where $\rho(t)$ is the density matrix representation of the evolved quantum state at time $t$, $\hbar$ is the reduced Planck constant, and $[a,b]=ab-ba$ is the commutator. In this work, we assume that the Hamiltonian is time independent; however, our method can be generalized to a time-dependent Hamiltonian $\H=\H(t)$. The initial state for the system is usually assumed to be pure, $\rho (0) = |\psi_{0}\rangle \langle \psi_{0}|$.

In general, the dynamics described in Eq.~\eqref{eq:von_neumann} cannot be classically solved exactly for systems with high dimensionality because the size of the Hilbert space grows exponentially. Fortunately, appropriate transformations such as the Jordan-Wigner~\cite{JWtransf} combined with Suzuki-Trotter formulas~\cite{suzukitrotter} allow for an efficient simulation of such Hamiltonians using quantum computers~\cite{Fauseweh2024}. In fact, simulating many-body quantum systems is a promising candidate for quantum advantage~\cite{lloyd1996,Fauseweh2024,lanes2025,Preskill2025}, but the main obstacle for these simulations to be accurate is the existence of quantum noise arising from qubit to environment coupling, crosstalk, imperfect gate implementations or measurement errors among others~\cite{9201447,Fowler_2012}. Assuming that the quantum system is subjected to Markovian noise, the dynamics of the open quantum system can be described by the \textit{Gorini-Kossakowski-Sudarshan-Lindblad (GKSL)} master equation~\cite{Lindblad1976,GKSLindblad}:
\begin{equation}\label{Eq: LindbladDE}
\frac{\partial}{\partial t}\rho_{\lambda}(t) = - \frac{i}{\hbar}[\H, \rho_{\lambda}(t)] +  \lambda \mathcal{L}(\rho_{\lambda}(t)),
\end{equation}
where $\mathcal{L}(\cdot) = \sum_w L_w\cdot L_w^\dagger - \frac{1}{2}\{ L_w^\dagger L_w,\cdot\}$ is the dissipation superoperator (Lindbladian) constructed by the jump operators $L_w$ that describe the interactions of the quantum system with its environment; the coefficient $\lambda\geq0$ is a coupling constant that defines the error rate, and $\{a,b\}=ab+ba$ is the anticommutator. The initial state of the system is the same as for Eq.~\eqref{eq:von_neumann}, $\rho_{\lambda}(0)=\rho(0)$, but its evolution differs. As a result, the expectation value $\Tr{(O\rho_{\lambda}(t))}$ of an observable $O$ will be corrupted by the noise and deviate from its noise-free value, $\Tr{(O\rho_0(t))}$~\cite{Cai_2023}.

In this context, the aim of QEM is to reduce the noise induced bias by post-processing the outputs of a number of noisy QPU runs~\cite{Cai_2023}. Nevertheless, the effect of noise is usually traded off for an increased variance (sampling overhead). ZNE and other related QEM techniques rely on the idea that, while the overall hardware noise of a quantum device cannot be reduced without QEC, it can actually be amplified. The noise level of the QPU is commonly quantified by means of a gain factor $G_k$, for which $G_1=1$ is the hardware limitation (base noise), \textit{i.e.} no amplification (See Appendix~\ref{Sec: Extrapo} and Refs~\cite{QEM2,UtilityIBM,commentZNE} for a formal definition of ZNE and noise amplification). Using different noise gains, $G_k$, one can obtain a set of noisy expectation values $\{\Tr{(O\rho_{G_k\lambda})}\}$, where $\lambda$ refers to the baseline error rate. For simplicity, we will abuse the notation  $\Tr{(O\rho_{G_k\lambda})}\equiv\Tr{(O\rho_{G_k})}$. Using such ensemble of noisy observables, error mitigation techniques aim to extrapolate to the zero-noise limit $\Tr{(O\rho_0)}$~\cite{ZNETemme}. In general, the main challenge of these extrapolation methods is that the decay of the expectation values as a function of the noise level is unknown (see Appendix \ref{Sec: Extrapo} for a detailed discussion), leading to biased estimations and high sampling overheads~\cite{Cai_2023}.

An extrapolation procedure can be formulated as a \emph{projection} problem. Indeed, suppose we are interested in recovering the expectation values of $N$ observables, $\{O_1,O_2,\ldots,O_N\}$ for a fixed time $\tau$ and can measure noisy expectation values at different noise gains $\{G_{j}\}_{j=1}^m$. Let $M\in\mathbb R^{N\times m}$ denote the measurement (design) matrix, obtained by stacking expectation values of the measured observables at different noise gains:
\begin{equation}\label{eq:measurement_matrix}
    \begin{aligned}
    M_{i,j} &:= \Tr(O_i\rho_{G_j}),
    & i&\in\{1,\ldots,N\},
      j\in\{1,\ldots,m\}
    \end{aligned}
\end{equation}
where $\rho_{G_j} \equiv \rho_{G_j}(\tau)$ denotes the quantum state at the fixed time $\tau$, evolved under the noise scaled by $G_j$.

Our goal is to obtain a set of vectors of optimal coefficients $\mathrm{x} = [x_{1}, \ldots, x_{m}]^{\top}$ such that their linear combination with the noisy expectation values provides the ideal noise-free values. This can be written as a least squares minimization problem:
\begin{equation}\label{projectionLSE}
    \hat{\mathrm{x}} = \arg \underset{\mathrm{x}\in\mathbb R^m}{\min} \lVert M\mathrm{x} - b\rVert_2^2,
\end{equation}
where $b\in\mathbb R^{N}$ is a target vector, \textit{i.e.} the vector containing the ideal values of the observables, stacked in the same order as the measurement matrix $M$:
\begin{equation}\label{eq:target_observables}
    b :=
        \begin{bmatrix}
            b_1 \\
            \vdots \\
            b_N
        \end{bmatrix},
        \quad
   \text{where }
    b_i = \Tr (O_i \rho_0).
\end{equation}
However, this problem is ill-posed since both the $b$ vector and the coefficients $\mathrm{x}$ are unknown. 

In the following section we present the GUESS method, which uses the decays of the Hamiltonian symmetries to learn the noise profile of the system and uses this information to mitigate the expectation value of the target observables. 

\section{GUiding Extrapolations from Symmetry decayS (GUESS)} \label{Sec: ProjectedLin}

Let us write an expression for the dynamics of the expectation value of an observable $O$ subject to noise. Projecting the GKSL Eq.~\eqref{Eq: LindbladDE} onto $O \in \mathcal{P}_n := \{ I, X, Y, Z \}^{\otimes n}$ and using the cyclic property of the trace, one obtains

\begin{align} \label{Eq: LindbladObs}
    \frac{\partial \langle O \rangle}{\partial t}
    =
    -\frac{i}{\hbar}\Tr\big([\H,O]\rho_\lambda(t)\big)
    +
    \Tr\!\left(\mathcal{L}^\dagger(O)\,\rho_\lambda(t)\right).
\end{align}

Here $\mathcal{L}^\dagger$ denotes the adjoint Lindbladian acting on observables, defined by the duality

\begin{equation}
    \Tr\left( O\,\mathcal{L}(\rho)\right)=\Tr\left(\mathcal{L}^\dagger(O)\,\rho\right)
\end{equation}

for all operators $O$ and density matrices $\rho$. If an observable is conserved under the unitary dynamics generated by the Hamiltonian, the unitary contribution vanishes identically. Consequently, the time evolution of its expectation value is governed solely by the dissipative term. In particular, for a symmetry operator $S$ satisfying $[\H,S]=0$, we find

\begin{equation} \label{Eq: LindbladSym}
    \frac{\partial \langle S \rangle}{\partial t} =  \Tr\left( \mathcal{L}^\dagger (S)\rho_\lambda(t)\right).
\end{equation}

Thus, while $\langle S\rangle$ is conserved under purely unitary evolution, any observed decay in its expectation value directly reflects the action of dissipative processes. This separation between unitary conservation and dissipative decay enables us to isolate noise-induced dynamics and to extract effective decay laws that directly characterize the QPU noise.

We therefore use this fact to first solve the least--squares problem~\eqref{projectionLSE} for a set of symmetry operators $\{S_\ell\}$, for which the corresponding expectation values $b_S$ are fixed under unitary evolution:
\begin{equation}\label{projectionLSE_symmetry}
    \hat{\mathrm{x}}_{\text{GUESS}} = \underset{\mathrm{x}}{\arg\min} \lVert M_{\mathrm S}\mathrm{x} - b_{\mathrm S}\rVert_2^2 \quad \text{  s.t. }\lVert\mathrm{x}\rVert_1=1,
\end{equation}
where $M_{\mathrm S}$ is the measurement matrix containing the noisy expectation values $\Tr{(S_{\ell}\rho_{G_j})}$ of the symmetry operators, stacked as in the matrix~\eqref{eq:measurement_matrix}. Here, we impose the restriction over the $1$-norm of the coefficients, $\lVert\mathrm{x}\rVert_1=1$, following what happens in Richardson extrapolation~\cite{ZNETemme} to avoid potential indeterminate systems of equations. Since the ideal noise-free expectation values of the operators $\{S_\ell\}$ are known exactly at any time $\tau$, the constrained least-squares problem~\eqref{projectionLSE_symmetry} can be solved, and we obtain the optimal coefficients $\hat{x}_{\text{GUESS}}$. Then, assuming that the noise action on the target observables $\{O_i\}$ and symmetry operators $\{S_\ell\}$ is similar, we can approximate the target vector~\eqref{eq:target_observables} as $b \approx M\hat{x}_{\text{GUESS}}$. It should be noted that in this learning scheme the weights of $\{S_\ell\}$ and $\{O_i\}$ must be consistent, since observables with different weights exhibit different decays (See discussion in Appendix~\ref{Sec: LindbladMesol}).

The GUESS mitigation method is summarized in Algorithm~\ref{alg: mitigation} and depicted in Figure~\ref{fig: Schematic}. Note that this method does not depend explicitly on the noise amplification factors, $G_j$. As long as the same amplification technique is used for both the symmetry and the target observable, the imperfections should have no effect, making GUESS robust to uncertainties on the noise scale (See Appendix~\ref{Sec: Extrapo}) and highlighting its potential utility in the early fault-tolerant regime, as discussed later. Finally, we consider two variants of this pipeline: the first one, described in Algorithm~\ref{alg: mitigation}, is referred to as linear GUESS. The second one, which we call exponential GUESS, is solved in the logarithmic domain and then the coefficients are exponentiated to obtain the final solution.

\begin{algorithm}[t!]
\caption{GUiding Extrapolations from Symmetry decayS (GUESS) QEM algorithm}
\label{alg: mitigation}
\begin{algorithmic}[1]
\Require Target Observables $\{\Tr(O_i\rho_{G_j})\}$, Measured Symmetries $\{\Tr(S_\ell\rho_{G_j})\}$
\Ensure Mitigated Observables $b_i$

    \State \textbf{procedure} \raggedright\textsc{ErrorMitigationProtocol}($\{\Tr(S_\ell\rho_{G_j})\}, \{\Tr(O_i\rho_{G_j})\}$)
    
    \Statex \raggedright $\triangleright$ \textbf{Step 1: Learning Parameters from Symmetries}
    \Function{SolveConstraints}{$\{\Tr(S_\ell\rho_{G_j})\}$}
        \State Define the matrix $M_S$ based on symmetry expectations values $\Tr(S_\ell\rho_{G_j})$
        \State Solve the least-squares problem to get $\hat{x}_{\text{GUESS}}$
        \State \Return $\hat{x}_{\text{GUESS}}$
    \EndFunction
    \Statex
    \Statex \raggedright $\triangleright$ \textbf{Step 2: Mitigating Target Observables}
    \Function{ApplyMitigation}{$\Tr(O_i\rho_{G_j}), \hat{x}_{\text{GUESS}}$}
        \State Define the matrix $M_{i,j}$ based on observable expectations values $\Tr(O_i\rho_{G_j})$
        \State Get the mitigated expectation values $b_i$ using $M_{i,j}$ and coefficients $\hat{x}_{\text{GUESS}}$
        \State \Return $b_i$
    \EndFunction

    \Statex
    \State \textbf{Main}
    \State $\hat{x}_{\text{GUESS}} \gets \Call{SolveConstraints}{\{\Tr(S_\ell\rho_{G_j})\}}$
    \For{each observable $O_i \in \{M_{i,j}\}$}
        \State $b_i \gets \Call{ApplyMitigation}{O_i, \hat{x}_{\text{GUESS}}}$
        \State $\text{mitigated\_results.append}(b_i)$
    \EndFor
    \State \Return mitigated\_results
\end{algorithmic}
\end{algorithm}

Extrapolation methods involve extending an assumed model beyond the range for which it has been directly supported by empirical evidence. While this allows us to mitigate noisy expectation values, it comes at the cost of increased uncertainty in the resulting value. From a statistical perspective, extrapolation typically leads to an increased variance~\cite{Cai_2023}. Studying the variance is important as it relates with the sampling overhead of the QEM method, \textit{i.e.} the amount of extra samples required to reach the same variance as the machine measurements. Parameter estimates used in the model are themselves subject to uncertainty, and their propagated effect grows outside the data-supported region. This is a direct consequence of error propagation where small uncertainties in estimated parameters can produce disproportionately large deviations in the extrapolated predictions~\cite{StatLearning}.

GUESS is composed of two different steps, in the first one we learn the coefficients $\mathrm{x}$ from a set of noisy expectation values of a set of symmetries, which has an associated variance. This variance is propagated to the coefficients by

\begin{align}
    \textrm{Cov}(\vec{\textrm{x}})_{i,\ell} = \sum_{j,k} J_{i, (j,k)} \sigma^2_{j,k} J_{\ell, (j,k)},
\end{align}
where $\sigma^2_{j,k}$ are the diagonal terms of the covariance matrix of the noisy expectation values and $J_{\ell, (j,k)}$ is the Jacobian defined by $J_{i, (j,k)} = \frac{\partial \textrm{x}_i}{\partial M_{s;j,k}}$. In the second step, we use the parameters to mitigate the target observable, which results in the final variance:

\begin{align} \label{Eq: varGUESS}
    \textrm{Var}(b_i)= \sum_j M_{i,j}^2\sigma_{x_j}^2  + \sum_j x_j^2 \sigma_{M_{i,j}}^2 + \sum_j \sigma_{x_j}^2 \sigma_{M_{i,j}}^2,
\end{align}
where $\sigma_{x_j}^2$ and $\sigma_{M_{i,j}}^2$ are the variances of the coefficients and the expectation values, respectively. The first and the second terms account for the contribution of the variance of $x_j$ and $M_{i,j}$ to the uncertainty of the noiseless expectation values. The last term in the expression accounts for the overlap of both uncertainties. For the variance of the exponential model and extended details on how we reach these expressions see Appendix~\ref{Sec: GUESS variance}.

\section{Practical Implementation via Hamiltonian impurities} \label{Sec: Simulations}

The GUESS method learns the decay of Hamiltonian symmetries as a function of the noise to mitigate target observables effectively. However, as discussed in Appendix~\ref{Sec: LindbladMesol}, the decay of an observable strongly depends on its weight. Theoretically, one can rescale the evolution time at which the symmetry is measured in order to take such effect into account. However, this is infeasible in practice because the time evolution is implemented via a quantum circuit based on the Trotter–Suzuki decomposition and such rescaling would significantly increase the circuit depth. While there are some important Hamiltonians with low weight symmetries such as Lattice Gauge theory Hamiltonians~\cite{LatticeGauge,cobos}, many important Hamiltonian classes of broad importance in quantum many body dynamics present small numbers of symmetries with very high weights~\cite{UtilityIBM,Fauseweh2024,PhysRevX.5.021027}. Furthermore, the noise in quantum devices is extremely heterogeneous or non-uniform~\cite{nonuniform} and, thus, the decay learned from a symmetry involving certain set of qubits may not be valid for another set of those.

To overcome these challenges, we introduce an impurity to the target Hamiltonian in order to have a suitable symmetry from which the noise can be learned correctly. Here, we define an impurity, $\mathcal{I}_{\bar{q}}$, as a local Hamiltonian perturbation which enforces a symmetry involving the set of qubits $\bar{q}$ of interest. Therefore, to mitigate a target observable, $O_w$, with arbitrary Pauli weight, $w$, which involves the measurement of $w$ qubits in $\bar{q}$, we consider the perturbed Hamiltonian $\H_I$:

\begin{align} \label{Eq: ImpHamil}
    \H_{\mathcal{I}}= \H_0 + \mathcal{I}_{\bar{q}},
\end{align}
where $\H_0$ refers to the target Hamiltonian. Crucially, the perturbed Hamiltonian fulfills the condition $[\H_{\mathcal{I}},O_w]=0$. 

By measuring the expectation value of the symmetry $O_w$ in the perturbed Hamiltonian, we can run the GUESS protocol described in Algorithm~\ref{alg: mitigation} to learn the optimal coefficients $\hat{\mathrm{x}}$, taking into account the observable weight and device noise non-uniformity. Here, it is critical to change the target Trotter circuit minimally so that error propagation is maintained, since circuit structure and error propagation play a pivotal role in the decay of the expectation values (See Appendix~\ref{Sec: Extrapo}). Even though mirror circuits and Clifford circuits have been ubiquitous in other error mitigation techniques such as Operator Decoherence Renormalization (ODR)~\cite{odr,cobos} or Clifford Data Regression (CDR)~\cite{CDR}, we instead preserve the structure of the target circuit as much as possible to capture potential non-trivial error propagation effects. 

To minimize changes in the circuit noise model when introducing the impurity, we maintain the number of two-qubit gates and their structure, ensuring that the circuit two-qubit depth remains unchanged. Crucially, the error rate of two-qubit gates is approximately one order of magnitude larger than the single-qubit gate error, making them the dominant source of noise and the primary target for amplifications and error mitigation in our approach~\cite{UtilityIBM,commentZNE}. Regarding error propagation, Pauli-twirling transforms noise channels into Pauli noise channels~\cite{Twirl2}, \textit{i.e.}
\begin{align} \label{Eq: Channel}
    \mathcal{N}(\rho) = \left( 1 - p\right) \rho +  \sum_{P \in \mathcal{P}^{\otimes n}} p_p P \rho P^\dagger,
\end{align}
where $P$ are Pauli operators in the $n$-fold Pauli group $\mathcal{P}^{\otimes n}$, $p_P$ are the probabilities associated to the Pauli operators $P$ and $p=\sum_{P \in \mathcal{P}^{\otimes n}} p_p$ is the probability of error.
It should be noted that the perturbation \eqref{Eq: ImpHamil} does not change the   noise channel propagation \eqref{Eq: Channel} significantly. By propagation here we refer to the structure of the noise channel after passing through a single quantum gate. Indeed, as we discuss in Appendix~\ref{Sec: Depol}, the noise channel propagation in the perturbed circuit remains close to that in the original circuit in diamond norm~\cite{diamondnorm}. This indicates that the perturbation accurately approximates the noise structure of the target circuit and, thus, the GUESS coefficients are accurate to mitigate the observables measured using the target circuits. This is confirmed with the experiments below.

\begin{figure*}[ht]
    \begin{overpic}[width=0.49\textwidth]{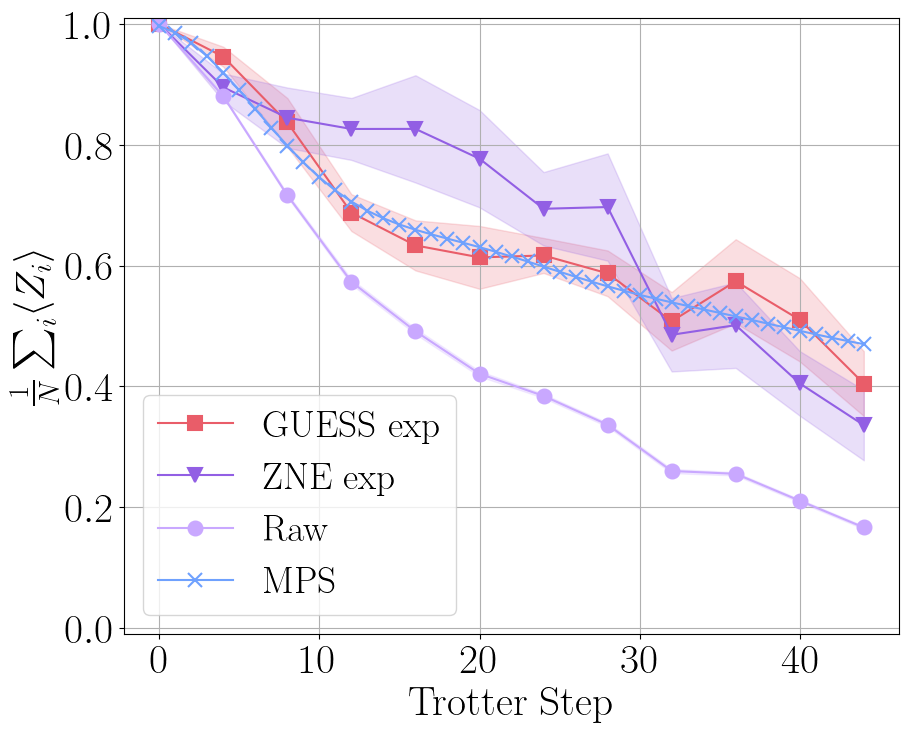}\put(1,80){\mbox{\large $(a)$}}
    \end{overpic}\hfill
    \begin{overpic}[width=0.49\textwidth]{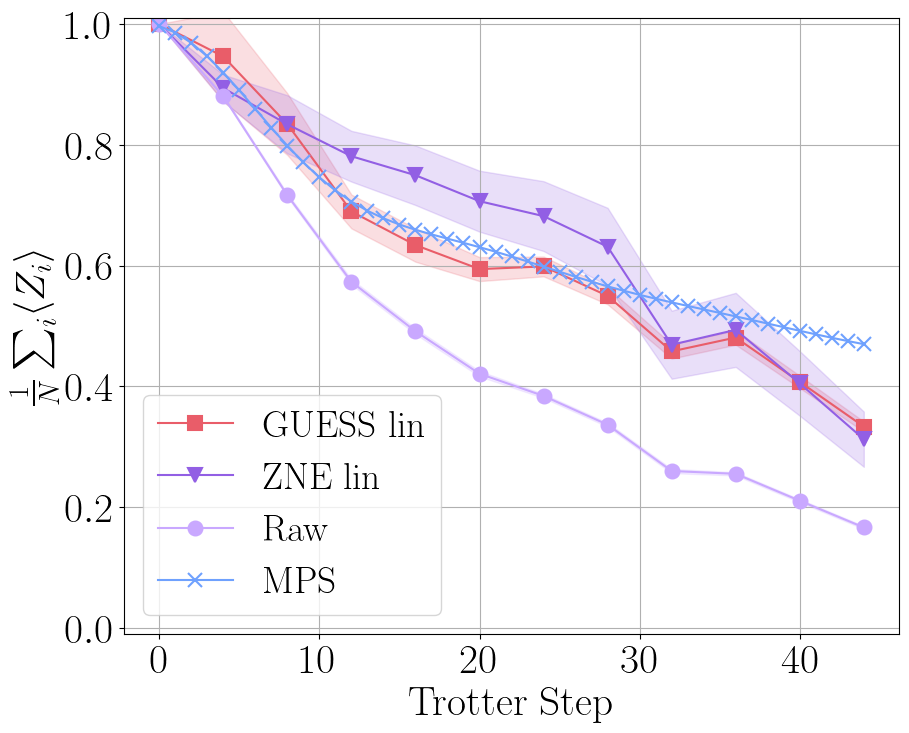}\put(1,80){\mbox{\large $(b)$}}
    \end{overpic}
    \medskip
    \medskip
    \begin{overpic}[width=0.49\textwidth]{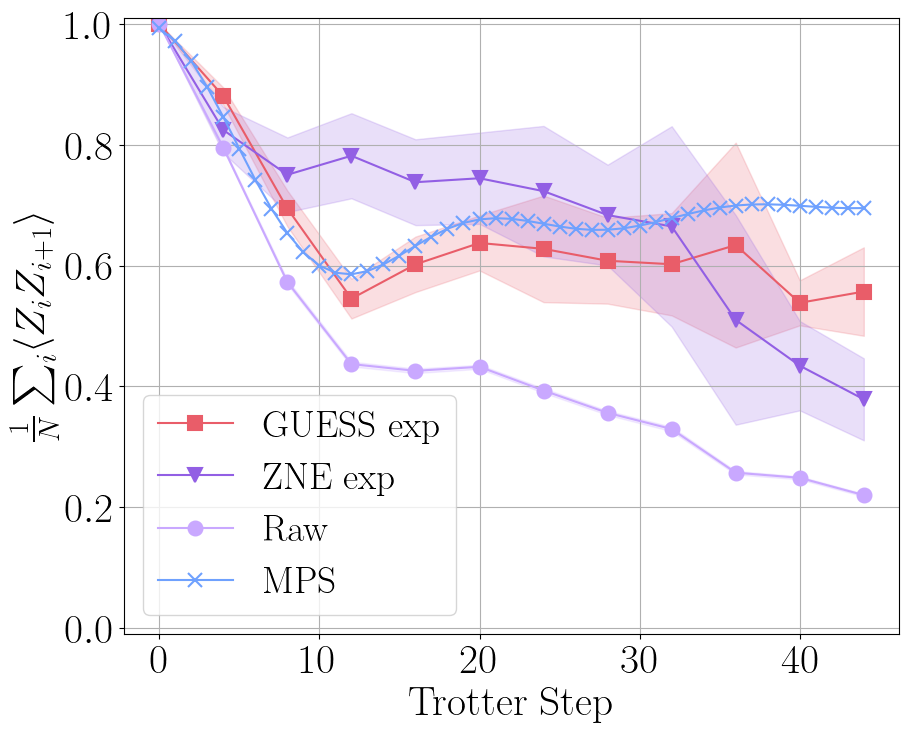}\put(1,80){{\mbox{\large $(c)$}}}
    \end{overpic}\hfill
    \begin{overpic}[width=0.49\textwidth]{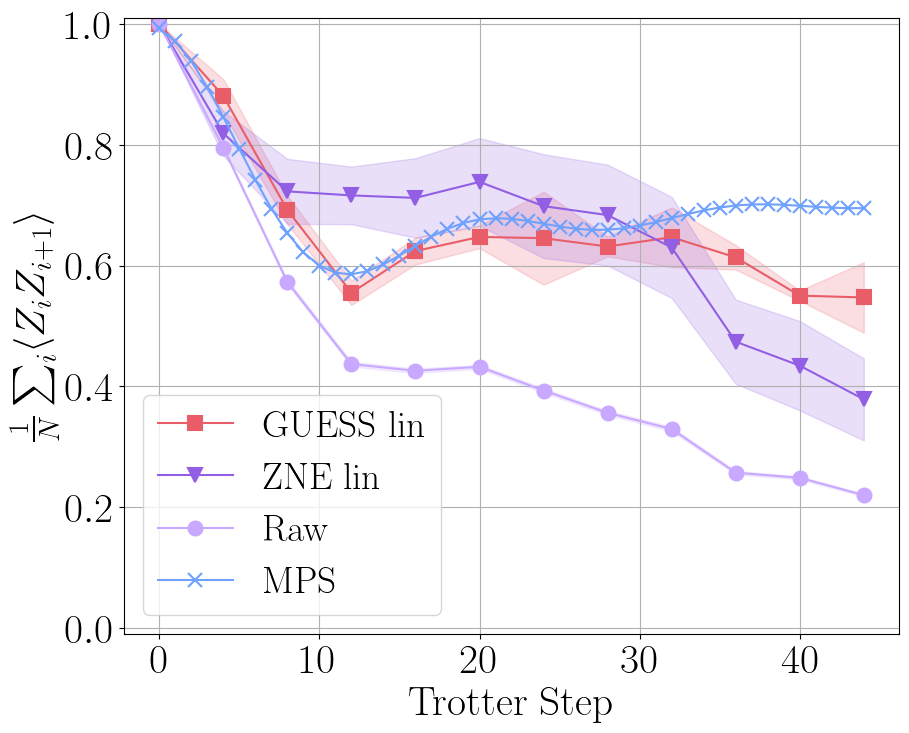}\put(1,80){{\mbox{\large $(d)$}}}
    \end{overpic}
    \caption{Simulation of the transverse field Ising chain with open boundaries, up to $44$ trotter steps and measuring each $4$ steps. The maximum two-qubit depth of the reported circuits is $176$, with a total number of $8712$ CZ gates for the baseline case without noise amplification $G=1$. We show the result for the best $20$ observables based on our statistical postprocessing method (see Appendix~\ref{Sec: Postpro}). In panels (a) and (b) we show the dynamics of the average magnetization (weight-$1$ observables) and in panels (c) and (d) of the average nearest-neighbor correlators (weight-$2$ observables); using the exponential and the linear model , respectively. The blue points (cross) refer to the ideal values, obtained from the MPS simulation using a time evolving block decimation (TEBD) algorithm with bond dimension of $150$ and $dt=t/\#\textrm{Trotter}$; the red (square) points to the extrapolated values using GUESS; and the dark (triangle) and light (circle) to ZNE and the raw machine measurements for $100$k shots, respectively.}
    \label{fig: Ising-wt1}
\end{figure*}

\begin{figure*}[ht]
\begin{overpic}[width=0.49\textwidth]{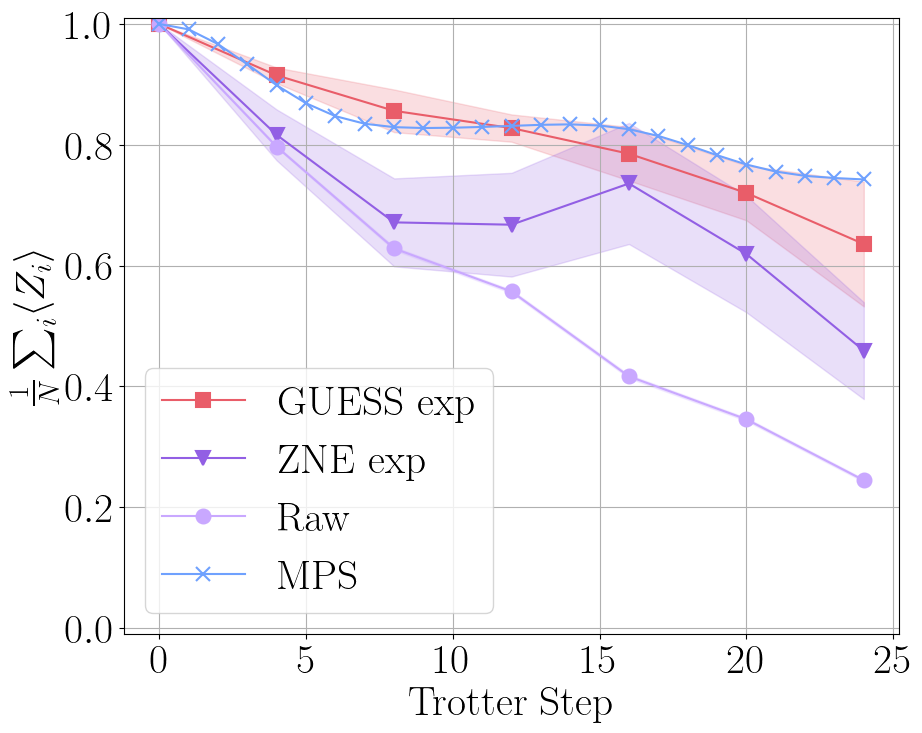}\put(2,80){\mbox{\large $(a)$}}
    \end{overpic}\hfill
    \begin{overpic}[width=0.49\textwidth]{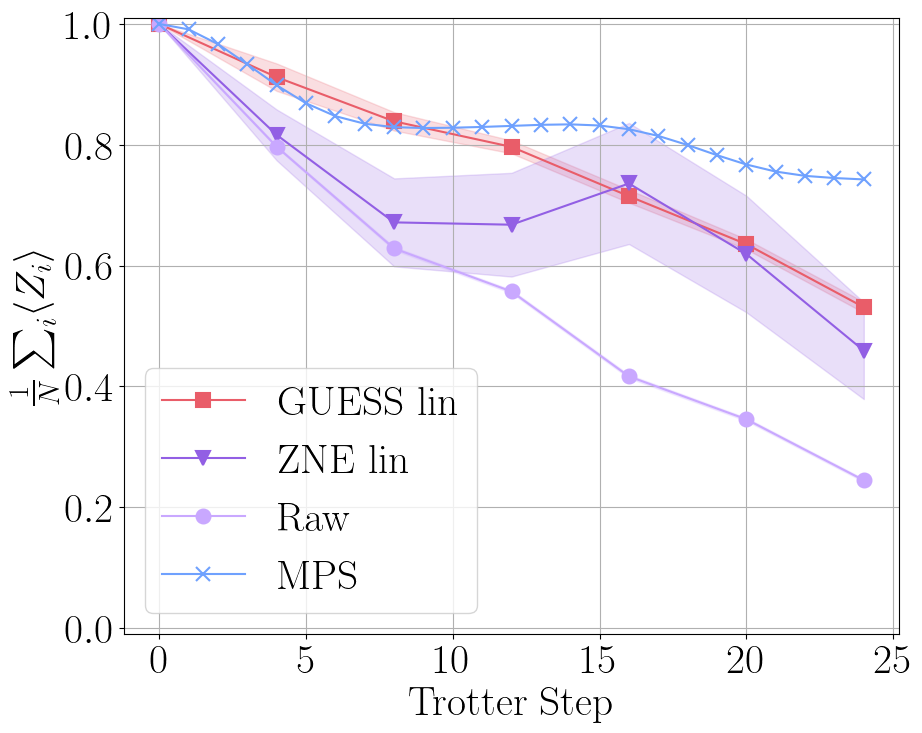}\put(2,80){\mbox{\large $(b)$}}
    \end{overpic}
    \caption{Simulation of transverse field $XZ$ Heisenberg chain with open boundaries, up to $24$ trotter steps measuring each $4$ steps. The maximum two-qubit depth of the reported circuits is $192$, with a total number of $9504$ CZ gates for the baseline case without noise amplification $G=1$. We show the result for the best $20$ observables based on our statistical postprocessing method (see Appendix~\ref{Sec: Postpro}). We show the averaged magnetization for (a) the exponential and (b) the linear models, where the blue points (cross) refer to the ideal values, obtained from the MPS simulation using a TEBD algorithm with bond dimension of $150$; the red (square) points to the extrapolated values using GUESS; and the dark (triangle) and light (circle) to ZNE and the raw machine measurements for $100$k shots, respectively.}
    \label{fig: Heis}
\end{figure*}

\section{Experimental validation on quantum hardware} \label{Sec: ExperimentResults}

We test the proposed GUESS mitigation technique on the '\texttt{ibm\_basquecountry}' QPU, an IBM Heron r2 processor consisting of $156$ superconducting transmon qubits in a HeavyHex architecture. Specifically, we simulate the time-evolution of a transverse field Ising model (TFIM) and an $XZ$ Heisenberg model in 1D, both with open boundary conditions. These quantum many-body Hamiltonians are extremely important testbeds to understand complex quantum phenomena~\cite{Fauseweh2024} and many problems are actually mapped to these classes of Hamiltonians by means of a Jordan-Wigner transformation~\cite{CerveraLierta2018exactisingmodel}. 

\subsection{Setup}

The corresponding Hamiltonians read as:

\begin{align}
    \H_\text{Ising} &= J \sum_{(i, j)} Z_{i}Z_{j}  +  h_X \sum_{i = 1}^{n} X_{i} \quad\text{    and    }  \label{eq: IsingHamil} \\
    \H_\text{Heis}&=   \sum_{(i, j)} \left(J_X X_{i}X_{j} + J_Z Z_{i}Z_{j}\right) +  h_X \sum_{i = 1}^{n} X_{i}. \label{eq: HeisHamil}
\end{align}
Here $\H_\text{Ising}$ corresponds to the TFIM with the coupling strength, $J$, and the field, $h_X$. The Hamiltonian $\H_\text{Heis}$ describes $XZ$ Heisenberg chain with a transverse field $h_X$; $J_X$ and $J_Z$ are the couplings in $X$ and $Z$ directions respectively. Here $X$, $Y$, and $Z$ are Pauli matrices. In the following, we have used $h_X=0.75$ and $J=1$ for $\H_\text{Ising}$; and $h_X=0.5$, $J_X=0.5$, and $J_Z=2$ for $\H_\text{Heis}$. For the initial state we have considered $\rho(0) = \ket{0}^{\otimes n}$.

We analyze the time dynamics of local observables for these models on a 1D lattice with $n=100$ sites and open boundary conditions. In 1D, these models can be efficiently simulated classically using Matrix Product State (MPS) tensor network techniques, which we employ as a benchmark. For time evolution, we use a first-order Trotter formula for $t=5$ ($44$ Trotter steps) and $t=3.75$ ($24$ Trotter steps) for the Ising and the Heisenberg models, respectively (See Appendix~\ref{Sec: Trotter} for details). To decouple algorithmic and hardware effects, our MPS-based tensor-network benchmarks use the same first-order Trotterization, so Trotter error does not enter the reported hardware-versus-simulation comparison. This set up enables a direct unrefined comparison between ZNE and GUESS, intentionally forgoing superior alternatives to isolate the core performance of both methods. Our goal is to demonstrate that the GUESS method can reliably capture decay behavior even when error amplification is imperfect. To this end, we adopt a gate-folding strategy for error amplification with noise factors $(1,1.2,1.5)$, which introduces heuristic uncertainties and creates a particularly challenging extrapolation scenario, compared with other amplification techniques such as probabilistic error amplification (PEA), which requires a costly noise‑learning stage~\cite{UtilityIBM} (See Appendix~\ref{Sec: Extrapo} for details regarding noise amplification). A detailed description of the quantum processor we used and additional information about noise mitigation techniques can be found in Appendix~\ref{Sec: IBM QPU}.

Additionally, both Hamiltonians considered exhibit a single global symmetry $S=X^{\otimes n}$ which involves all the spin sites. Here, we neglect the total energy as a symmetry, since it would require the measurement of numerous low-weight observables. Therefore, we need to employ the impurity method proposed above in order to run the GUESS mitigation protocol for local observables, making this simulation scenario an ideal setting to evaluate the performance of the GUESS method.

\subsection{Results}

We depict the recovered dynamics of the Ising model, $\H_\text{Ising}$, for $20$ observables of weight $1$ (average magnetization) and $20$ observables of weight $2$ (two-point correlation) in panels (a) and (b) of Fig.~\ref{fig: Ising-wt1}, respectively. The  blue (cross) points are the ideal values obtained by means of MPS simulations (see Appendix~\ref{Sec: ClassicalSim}); the red (square) points are the extrapolated values using GUESS; and the dark (triangle) and light (circle) purple refer to our implementation of ZNE and the noisy raw signal ($G=1$) measured using the QPU, respectively. Here, we enforce impurities $\mathcal{I}_i=-h_IX_i$ and $\mathcal{I}_{i,i+1}=-\sum_{j\in\{i,i+1\}}h_IX_j$ for weight $1$ and $2$ observables, respectively; which preserve the number of two-qubit gates. These terms are designed to enforce the commutation relations $[\H_\mathcal{I},Z_i] = 0$ and $[\H_\mathcal{I},Z_iZ_{i+1}] = 0$, respectively. Furthermore, we show the results for the dynamics of the average magnetization (20 weight $1$ observables) in the considered Heisenberg model~\eqref{eq: HeisHamil}, in Fig.~\ref{fig: Heis}. In this case, enforcing symmetries involves modifying the $J_{X}X_{i}X_{i+1}$ interaction terms as well as single-qubit transverse field terms. Thus, we introduce a set of symmetry preserving two-qubit gates, specifically $J_{X}Z_{i}Z_{i+1}$ and $J_{X}Z_{i-1}Z_{i}$ interactions, to maintain the two-qubit gate count. The impurity required to enforce symmetry on the $i^{\text{th}}$ qubit is defined as follows:

\begin{align}
    \mathcal{I}_{i}= -h_H X_i - J_{X}X_{i-1}X_{i} - J_{X}X_{i}X_{i+1} +J_{X}Z_{i-1}Z_{i} + J_{X}Z_{i}Z_{i+1}.
\end{align}

The choice of these impurities has a negligible impact on how the noise propagates (see Appendix~\ref{Sec: Depol} for details). The results in Fig.~\ref{fig: Ising-wt1} and Fig.~\ref{fig: Heis} clearly demonstrate that the GUESS method consistently produces expectation values with higher accuracy compared to ZNE. The resulting dynamics are consistent with the MPS ground truth for around $40$ Trotter steps in the Ising model case and $20$ Trotter steps in the $XZ$ Heisenberg model, for both the exponential and the linear versions of GUESS. The relative error achieved by GUESS for these time windows is around $10\%$ (See Figure~\ref{fig:placeholder} in Appendix~\ref{Sec: RelErr} for quantification of the relative error). Such a relative error is consistent with the accuracies generally accepted in current state‑of‑the‑art results obtained on quantum hardware~\cite{cobos,UtilityIBM,Kim2023,Mildenberger2025,Cochran2025,OTOC,HeisQMB,odr,mitCentral}. A close inspection reveals that the exponential GUESS is able to achieve lower error but presents a higher variance than the linear one, reflecting a trend similar to that observed in standard ZNE. Furthermore, both the raw signal and ZNE fail to capture the dynamics of the observables, showing relative errors beyond $20\%$ for early-time dynamics. Note that the low relative error observed for ZNE occurs coincidentally, as the noisy signal crosses the true value due to decay induced by noise.

The QEM methods used in this work yield mitigated values that have no physical meaning, i.e. $|\langle O \rangle|>1$, indicating that the obtained value is wrong. In Table~\ref{Table: Non-phys}  we show the percentage of non-physical results for all simulations, further details regarding their origin are provided in Appendix~\ref{Sec: Non-physical}. We leverage this information to switch the mitigation model between the two variants (linear and exponential) for both ZNE and GUESS. Specifically, if an overshoot occurs, the exponential method is downgraded to the linear variant, and conversely for the linear case. When both methods yield non-physical results, we default to the raw measurement. The reported values in Figures~\ref{fig: Ising-wt1} and~\ref{fig: Heis} correspond to the averages over $20$ selected observables. As detailed in Appendix~\ref{Sec: Postpro}, we initially measure $40$ observables and then post-process the $20$ most reliable ones based on the expectation values and their variance from the symmetry. Since the ideal value of the conserved observable is known, we can make use of it to determine outliers based on expectation values and standard deviations of symmetry machine measurements. This allows us to select the least noisy qubits at the time of measurement. Moreover, for systems where all observables are physically relevant, this technique can be used to identify the measurements that need to be performed again due to unreliable baseline values. Appendix~\ref{Sec: Postpro} provides a detailed explanation and includes unprocessed results for all measured $40$ observables. Leveraging observable uncertainties and discarding mitigation outcomes that violate physical constraints is a widely adopted practice in the literature~\cite{UtilityIBM}.

\begin{table}[ht]
\centering
\caption{The proportion of non physical values, values above $1$ or below $-1$, for different Hamiltonians and QEM methods. It shows the average percentage of non physical values of the $40$ observables along all the trotter steps.}
\label{Table: Non-phys}
\begin{tabular}{|c||c|c|c|c|}
\hline
 \% & ZNE Lin & ZNE exp & GUESS Lin & GUESS exp  \\
\hline
Ising wt(1) & $11.59$ & $81.14$  & $0.07$ & $2.05$  \\\hline
Ising wt(2) & $9.77$ & $89.54$  & $7.5$ & $6.36$ \\\hline
Heisenberg & $24.17$ & $96.67$ & $2.92$  & $6.67$  \\
\hline
\end{tabular}
\end{table}

\section{Discussion and Conclusion} \label{Sec: Discussion}
In this paper, we have proposed a QEM technique, GUiding Extrapolations from Symmetry decayS (GUESS), which improves ZNE by leveraging prior knowledge of the functional dependence of observable decay on noise strength, thereby reducing both estimation bias and sampling overhead. Additionally, our method addresses important uncertainties that arise in real hardware due to non-ideal amplification techniques, which lead to inaccuracies in the effective noise amplification factors.

The core principle behind GUESS is that expectation values of Hamiltonian symmetries are conserved through the evolution of quantum many-body systems. Specifically, we introduce a Hamiltonian impurity construction that minimally perturbs the target problem to introduce a symmetry of interest matching the weight and the involved qubits of the target observable. The symmetry of the perturbed Hamiltonian enables us to learn optimal coefficients that are then used to mitigate the target observables of the original (unperturbed) Hamiltonian to their noiseless values. We discuss the conditions under which the Trotter quantum circuit for the perturbed Hamiltonian accurately captures the noise propagation of the target circuit. In particular, we show that the learned noise behavior is robust when the impurity does not have a hard impact on the noise propagation and when the two-qubit gate count (circuit depth) is preserved. This is crucial on current quantum hardware, where noise is predominantly limited by two-qubit operations and where those are the primary source targeted by error mitigation techniques.

We demonstrate the practicality of GUESS by simulating the dynamics of average magnetization and nearest-neighbor two-body correlators for a 1D transverse-field Ising model and an $XZ$ Heisenberg model at utility scale (100 qubits) on the IBM Heron QPU \texttt{ibm\_basquecountry}. When Pauli Twirling is enabled, noise in superconducting qubits is well described by Pauli channels, and the error propagation is similar for the target and impurity gate sequences. Crucially, our method recovers noisy quantum observables with $\sim10\%$ error for up to $~40$ and $~20$ Trotter steps for the Ising and Heisenberg models, respectively. We also observe lower standard deviation, indicating a lower sampling overhead. We employ a first-order Trotter formula, where each two-body interaction is transpiled to two entangling CZ gates. In this setting, GUESS achieves accurate mitigation for circuits with depths up to $160$ and $\sim 8000$ two-qubit gates.

Our approach relates to methods that learn circuit noise to mitigate noisy expectation values. Clifford Data Regression (CDR)~\cite{CDR} replaces the target circuit with one largely composed of Clifford gates to fit a linear ansatz between ideal and noisy data. However, noise propagation differs significantly between Clifford and non-Clifford circuits~\cite{UtilityIBM,caimultiexp}, which can cause learned corrections to deviate from the target. Furthermore, CDR methods are expensive in terms of hardware use as they require large training sets consisting of hundreds of circuits per observable \cite{CDR,CDR2}. Operator decoherence renormalization (ODR)~\cite{odr} typically uses mirror circuits to infer decay factors of observables; while often effective heuristically, mirrored circuits can also alter the underlying noise behavior, limiting the guarantees of transfer. GUESS with Hamiltonian impurities is designed to address these discrepancies so that the learned decay resembles the target circuit as closely as possible. Notably, when a single noise factor suffices, GUESS reduces to an ODR-like protocol that uses a perturbed circuit rather than mirror circuits. As such, methods like CDR, ODR, and machine-learning-based QEM~\cite{CDR,odr,Liao2024,CDR2} can adopt GUESS to learn noise behavior more accurately. The quantum overhead of GUESS matches ODR, that is, two circuits per observable (target and perturbed).

In our study, we use fractional gate folding as the noise amplification protocol. As discussed in Appendix~\ref{Sec: Extrapo}, such a noise amplification technique is heuristic and can be inaccurate~\cite{gatefolding-imp}. Realistic noise can be nonlinear or change its figure at large fold factors, causing extrapolations to diverge from the ideal results due to inaccurate scaling~\cite{gatefolding-prob}. However, we deliberately tested our learning method over the most demanding conditions and, as shown in the results, it remains reliable despite imperfect amplification. Notably, the imprecise performance of baseline ZNE observed here stems from the mismatch between the scaling factors considered and the gate-folding implementation. Following this line of reasoning, one may argue that ZNE would be effective if more precise amplification techniques, such as PEA~\cite{UtilityIBM}, were used. We note that these methods would equally benefit GUESS, extending the mitigation to longer times. In this work, our aim is to show that GUESS operates reliably with low quantum overhead, especially given that PEA requires a costly noise‑learning stage and detailed device‑level characterization.

Within this framework, we believe that GUESS can serve as a practical bridge for hybridizing ZNE-based QEM in the early fault-tolerant regime. Combining QEC with QEM is considered essential to enhance the ability of early fault-tolerant machines in providing accurate results~\cite{MegaquopMachine,logicalQEM1,logicalQEM2,logicalQEM3,jeon2026}. However, the current literature on the topic is limited and focused on theory. Experience with near-term hardware has demonstrated to the QEM community that theory and experiment frequently diverge, complicating practical realization. Moreover, logical gates are fundamentally different to the physical gates used in error mitigation, involving protocols such as lattice surgery, transversal gates, teleportation of distilled magic states or code deformation, among others~\cite{litinski,quera,MagicExpQuera,Horsman_2012}. Thus, imposing the requirement that a logical noise learning protocol accurately supports PEA-like amplification protocols over logical gates appears complex and costly. Additionally, mapping increased physical noise to increased logical noise remains challenging, and fully understanding logical noise propagation (especially with non-Clifford operations) may be challenging. Nevertheless, logical noise may be amplified without knowledge of the amplification factors, for example by reducing the distance of the QEC code~\cite{logicalQEM2}, folding the logical gates or intentionally inducing failures in parts of the error-detection process, i.e. tweaking the decoder~\cite{decoders}. This would make it possible to apply the GUESS protocol to a logical circuit and mitigate the resulting effect on target observables, in the same spirit as the experiments presented in this work. We expect that GUESS can find its niche in combining ZNE-based QEM with logical qubits, at least for early fault-tolerance era.

\section{Acknowledgements}
We would like to thank Antonio de Marti i Olius, Paul Schnabl, Reza Dastbasteh, Pedro M. Crespo, Jesus Cobos, Pedro Rivero and Sergiy Zhuk for many useful discussions and recommendations. We also acknowledge the IBM Quantum Algorithm Engineering team for their insights and contributions.

This work has been funded by the BasQ strategy of the Department of Science, Universities, and Innovation of the Basque Government through the ``Projection  Methods  for  Device  Error  Mitigation (ZNEProDEM)" and ``Extrapolation
of Von Neumann Dynamics beyond the reach of current Utility Scale Devices (VNDExUSD)'' joint research agreement projects. We acknowledge the use of IBM Quantum Credits for this work.

\bibliography{Bibliography}

\clearpage
\onecolumngrid
\appendix

\clearpage

\section{Zero noise extrapolation} \label{Sec: Extrapo}

Zero noise extrapolation (ZNE) is based on the conception that even if the noise in a circuit cannot be reduced (because there is no QEC available, for example), it can actually be amplified. Based on this idea, ZNE aims to recover the ideal expectation value of a target observable, $\Tr(O\rho_0)$, by means of a set of noisy expectation values$\{\Tr(O\rho_{\lambda_1}),\cdots,\Tr(O\rho_{\lambda_{n+1}})\}$, where by $\rho_{\lambda_i}$ we refer to the quantum state obtained after the noisy evolution with rate $\lambda_i$ and $O$ is the observable~\cite{Cai_2023,ZNETemme}. In practice, a quantum circuit presents several elements with distinct noise rates, but if all of them are amplified with the same factors, the situation is equivalent~\cite{commentZNE}. From the theory of Richardson extrapolation, it can be shown that the noisy observable can be approximated using $n$ noise rates as a polynomial of order $n$ over $\lambda$, resulting in the following approximation of the ideal value~\cite{ZNETemme,commentZNE}:
\begin{equation}\label{eq:estimate}
    \Tr(O\rho_0) \approx \sum_{j=1}^{n+1 }\gamma_j\Tr(O\rho_{\lambda_j}),
\end{equation}
where the coefficients, $\gamma_j$, result to be the Lagrange interpolators, \textit{i.e.} $\gamma_j = \prod_{m\neq j}\frac{\lambda_m}{\lambda_j-\lambda_m}$. 

Although the theory of Richardson extrapolation is mathematically sound, linear or exponential fits are usually employed in practice due to numerical instabilities arising from the variance of the measured observables and Runge's phenomenon~\cite{UtilityIBM}. Empirically, exponential extrapolation has shown to be the most accurate when the functions describing the decay behave in such a way. However, the existence of multi-exponential decays arising from noise propagation in non-Clifford circuits, \textit{i.e.} the ones of interests, makes the choice of the exponential extrapolation unstable, potentially providing unphysical values~\cite{caimultiexp,UtilityIBM}. Furthermore, this extrapolation becomes unstable for large circuit depths when the values start to be close to zero~\cite{UtilityIBM}. Linear extrapolation results to be more stable, but is not too accurate due to its conservative nature. This blindness to which curve to fit is one of the challenges in making ZNE more accurate and less costly~\cite{Cai_2023}.

Additionally, the noise amplification protocols are also a challenge for ZNE based mitigation methods. The first proposed noise amplification method was pulse stretching, based on modifying the pulse length of the quantum gates resulting in equivalent but noisier versions of those~\cite{ZNETemme,QEM2}. While accurate, this approach has been dismissed in practice due to requiring intensive pulse level control. Therefore, the most established noise amplification methods nowadays are gate folding and probabilistic error amplification~\cite{Giurgica_Tiron_2020,QEMbestpractices,UtilityIBM}. Gate folding is based on the idea of amplifying noise by means of repeated sequences of the gates and their conjugated ones, \textit{i.e.} $U\rightarrow U(U^\dagger U)^j $~\cite{Giurgica_Tiron_2020,QEMbestpractices}. One can easily see that folding the whole circuit leads to having access to $j \in \{1,3,5,...\}$ amplification factors, so partial folding was proposed as an heuristic method to access other amplification factors (generally non-integers) that are beneficial to ZNE~\cite{Giurgica_Tiron_2020,QEMbestpractices,chebyshev,UtilityIBM}. Partial folding is a cheap way to amplify noise, but unfortunately lacks exactness due to its heuristic nature. On the contrary, PEA is based on the noise learning concepts that were conceived for probabilistic error cancellation (PEC)~\cite{PEC3}. This technique relies on a correct modeling of the noise in a quantum circuit so that it can be efficiently learned and subsequently amplified by randomizing the circuit runs~\cite{UtilityIBM}. This technique is more accurate than gate folding, but requires accurate specification and learning of the noise model, as well as having to effectively tune the circuit randomizations. All this becomes more complicated considering noise drift~\cite{UtilityIBM,EtxezarretaMartinez2021} since the noise has to be learned almost every time that the circuits are executed.

ZNE methods have shown to be very sensitive to all the effects described, \textit{i.e.} unknown multi-exponential effects and imperfect noise amplifications, thus, proposing methods to improve the stability and accuracy of ZNE is fundamental for QEM.

\section{Lindblad Master equation with Pauli noise and observable weight considerations} \label{Sec: LindbladMesol}

The evolution of an open quantum system with Markovian noise can be described by the GKSL master equation defined in Eq.~\eqref{Eq: LindbladDE}. The evolution of the quantum state depends on the type of jump operators $L_w$ that describe the noise in the system. However, assuming that such general noise maps are Pauli twirled, we then assume that the jump operators are given by the Pauli matrices. For instance, the approximation here is enough for our purposes since we will actually Pauli twirl the noise when running experiments on real hardware. Finally, we will assume uncorrelated depolarizing noise affecting each of the qubits equally for the sake of simplicity here, but its generalization to Pauli noise containing higher weight terms is straightforward. Furthermore, a non-uniform noise model would result in having a specific coupling rate, $\lambda_w$, for each jump operator in Eq.~\eqref{Eq: LindbladDE}; the analysis here can also be generalized in an straightforward manner to such scenario \cite{commentZNE}. Under this model, the dissipator in the Lindblad equation reads as:

\begin{align}
    \lambda\mathcal{L}(\rho_\lambda(t)) &=\sum_{w=1}^{n} \lambda\left(L_w\rho_\lambda(t) L_w^\dagger - \frac12 \{L^\dagger_wL_w , \rho_\lambda(t)\}\right) \nonumber \\   
    &= \sum_{w=1}^n \lambda \left(X_w\rho_\lambda(t) X_w + Y_w \rho_\lambda(t) Y_w + Z_w \rho_\lambda(t) Z_w - 3 I^{\otimes n} \rho_\lambda(t) I^{\otimes n}\right),
\end{align}
where $n$ is the number of sites of the system and by $X_w,Y_w,Z_w$ in the second line we refer to a Pauli operator that only contains a single non-trivial operator at position $w$. Now, defining an arbitrary Pauli observable $O$ and taking the trace of Eq.~\eqref{Eq: LindbladDE} we get the following expression after some algebra:

\begin{align} \label{eq: TimeEv O}
    \frac{\partial \Tr{(O \rho_\lambda(t))}}{\partial t} = &-\frac{i}{\hbar} \Tr{([\H,O]\rho_\lambda(t))}  \nonumber\\
    &+ \Tr{\left( O \sum_{w=1}^n \lambda \left(X_w\rho_\lambda(t) X_w + Y_w \rho_\lambda(t) Y_w + Z_w \rho_\lambda Z_w - 3 I^{\otimes n} \rho_\lambda(t)I^{\otimes n}\right) \right)}.
\end{align}

To simplify the expression, we first focus on the jump operators and using the cyclic property of the trace we rewrite it as:

\begin{align}\label{eq: TimeEv O}
     \Tr\left(  \sum_{w=1}^n \lambda \left(X_w O X_w\rho_\lambda(t)  + Y_w O Y_w \rho_\lambda(t)  + Z_w O Z_w \rho_\lambda(t) - 3 O \rho_\lambda(t) \right) \right).
\end{align}

As said before, the observables are Pauli operators, so, due to their commutation relations, the following relationships are satisfied:

\begin{align}
    P_w P_v P_w = \begin{cases}
    P_v \text{ if } [P_w,P_v]=0 \\
    -P_v \text{ if } \{P_w,P_v\}=0 \\
    \end{cases}.
\end{align}

Thus, if the observable commutes with all the single-weight Pauli operators at position $w$, which is only possible if $O_w= I$, we get

\begin{align}
    \Tr\left( \lambda  \left(X_w O X_w  + Y_w O Y_w   + Z_w O Z_w   - 3 I  \right) \rho_\lambda(t) \right) = 0,
\end{align}
and, otherwise,
\begin{align}
     \Tr\left( \lambda  \left(X_w O X_w  + Y_w O Y_w   + Z_w O Z_w   - 3 O \right)\rho_\lambda(t)\right) =  \lambda\Tr\left(-4O \rho_\lambda(t)\right),
\end{align}
since if $O$ commutes with one of the single weight Pauli operators, it will anticommute with the rest at position $w$. Using these facts, we can rewrite Equation~\ref{eq: TimeEv O} as:

\begin{align} \label{eq: Dif.Eq same lambda}
    \frac{\partial Tr(O\rho_\lambda(t))}{\partial t} &= -\frac{i}{\hbar} \Tr([\H,O]\rho_\lambda(t)) + \Tr\left( - \lambda 4 \text{wt}(O) O \rho_\lambda(t) \right) \nonumber\\
    &= -\frac{i}{\hbar} \Tr([\H,O]\rho_\lambda(t)) - \lambda 4 \text{wt}(O) \Tr\left( O \rho_\lambda(t) \right),
\end{align}
where $\text{wt}(O)$ refers to the operator weight and arises because the observable can only anticommute at $\text{wt}(O)$ positions with the weight one Pauli operators. Renaming $\Tr(O\rho_\lambda(t))$ as $y$, we can rewrite the last equation as:
\begin{align}
    \frac{\partial y}{\partial t}&= -\frac{i}{\hbar} \Tr([\H,O]\rho_\lambda(t)) - \lambda 4 \text{wt}(O) y \quad \rightarrow  \quad
    \frac{\partial y}{\partial t} + p(t) y = f(t),
\end{align}
where $f_k(t)=-\frac{i}{\hbar} \Tr([\H,O]\rho_\lambda(t))$ and $p(t) = 4\lambda \text{wt}(O)$. The solution of this differential equation has the following form:

\begin{align}
    y(t) &= Ae^{P(t)} + v(t) e^{P(t)}, \\
    \text{where}& \begin{cases}
        P(t) = -\int_0^t p(t) dt = - 4 \lambda  \text{wt}(O) t, \\
        A = \Tr(O\rho_\lambda(0)), \\ 
        v(t) = \int f(t) e^{-P(t)} dT = \int_0^t \Tr([\H,O]\rho_\lambda(t)) e^{4 \lambda \text{wt}(O)t},
    \end{cases}
\end{align}

Rewriting the equation we get

\begin{align} \label{eq: DiffEq solution same lambda}
    \Tr(O\rho_\lambda(t)) = \left( \text{A} + \int_0^t \Tr([\H,O]\rho_\lambda(t)) e^{4 \lambda \text{wt}(O)t}\right) e^{-4 \lambda \text{wt}(O)t},
\end{align}
where $A$ is the initial condition ($Tr(O\rho_\lambda(0))$). From this expression, we can observe how the decay of an observable depends on the weight of its associated Pauli operator. Therefore, it is clear that the weight of an observable is something to take into account when aiming to learn the decay function of the expectation value of a quantum observable as a function of the noise factor.

\section{Classical Simulations} \label{Sec: ClassicalSim}

\subsection{MPS simulation}\label{Sec: MPS}

As explained in the main text, we tested our mitigation method for two different quantum many body Hamiltonians in 1D with open boundary conditions. This scenarios are perfect for matrix product state (MPS) based tensor network simulation of the systems, enabling us to use the resulting data as ground truth to evaluate the performance of the QEM techniques under study. We employed the well-known python package named ``TeNPy"~\cite{tenpy} and a time evolving block decimation (TEBD) algorithm in order to obtain the exact values of the considered observable evolutions. We implemented the time-evolution using the same first order Trotter formulas a time evolving block decimation (TEBD) algorithm for implementation in real hardware so that we can decouple the error from the Trotter error and, thus, uniquely study the error coming from quantum noise. For the sake of assuring MPS convergence, we simulated each Trotter step with varying bond dimensions to then check convergence. For the simulations performed, we reached up to $150$ bond dimension to assure convergence.

\section{IBM Quantum Computer Simulation} \label{Sec: IBM QPU}

\begin{figure*}[ht] 
\includegraphics[width=0.7\textwidth]{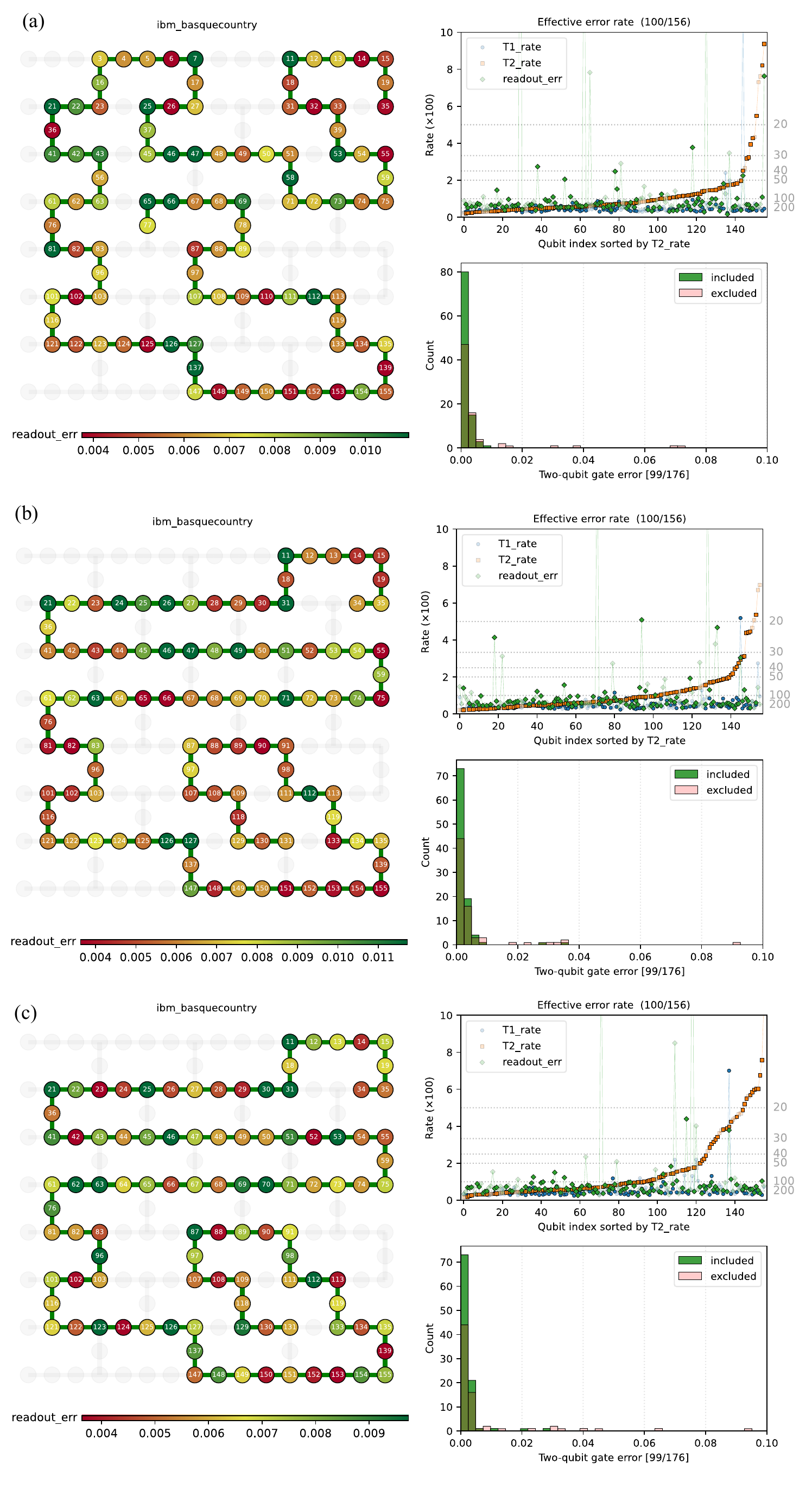}
    \caption{Selected qubit layouts and harwdware status of the '\texttt{ibm\_basquecountry}' at the times of circuit runs. (a) Transverse field Ising model experiments for magnetization observables, (b) Transverse field Ising model experiments for correlator observables, (c) Transverse field $XZ$ Heisenberg model experiments for magnetization observables.}
    \label{fig:HWstatus}
\end{figure*}

The hardware experiments were carried out on the IBM Heron r2 quantum processing unit '\texttt{ibm\_basquecountry}' consisting of $156$ superconducting transmon qubits in a heavy-hex layout. The device operates at $250$k circuit layer operations per second and, at the time of the circuit runs, the average two-qubit gate (CZ) error rates were $0.179 \%$, $0.191 \%$ and $ 0.185 \%$ for the simulation of the average magnetization and the nearest-neighbor correlators of the Ising model and the average magnetization of the $XZ$ Heisenberg model, respectively. We depict the specific status of the QPU and the selected qubit layout at the time of each circuit run in Figure~\ref{fig:HWstatus}.

The figures also showcase the selected qubit layout for each of the $100$ site 1D Hamiltonians implemented, which were selected based on the best qubits of the system in terms of readout error at that time. For each case, we measure $40$ observables to obtain the average magnetization and nearest neighbor correlators shown in the main text. As stated in the main text, we use fractional gate folding for noise amplification with $3$ noise factors $(1,1.2,1.5)$. Additional noise‑reduction techniques included Dynamical Decoupling (DD) to reduce crosstalk errors~\cite{DD1,DD2}, Twirled Readout Error eXtinction (TREX) to reduce the readout error~\cite{TREX} and Pauli Twirling to ensure stochastic Pauli noise~\cite{Twirl1,Twirl2}, with $32$ different randomizations and $3125$ shots per each one. Finally, we run $100$k shots for all circuit runs.

\subsection{Trotterization} \label{Sec: Trotter}

The implemented quantum circuit for the many-body Hamiltonians have been obtained by using the following first order Suzuki-Trotter formulas. Each Trotter step is then defined as:

\begin{align}
    e^{-i \textrm{H}_\text{Ising} \delta t} &\simeq  e^{H_{ZZ}^{\text{odd}}}e^{H_{ZZ}^{\text{even}}}e^{H_X } \\
    e^{-i \textrm{H}_\text{Heis} \delta t} &\simeq  e^{H_{XX}^{\text{odd}}}e^{H_{XX}^{\text{even}}}e^{H_{ZZ}^{\text{odd}}}e^{H_{ZZ}^{\text{even}}}e^{H_X},
\end{align}

where
\begin{align}
    \H_{\mathcal{P}\mathcal{P}}^{\text{odd}}&=-i J_\mathcal{P} \delta t\sum_{i = 1}^{\left \lfloor(N-1)/2\right \rfloor}  \mathcal{P}_{2i-1}\mathcal{P}_{2i} \\ \H_{\mathcal{P}\mathcal{P}}^{\text{odd}}&=-i  J_\mathcal{P} \delta t \sum_{i = 0}^{\left \lfloor(N-2)/2\right \rfloor}  \mathcal{P}_{2i}\mathcal{P}_{2i +1} \\
    \H_{X} &=-ih \delta t\sum_i  X_i .
\end{align}
where $\delta t = t /N_t$, being $N_t$ the number of trotter steps. We have used $N_t=44$ with $t=5$ for both simulations of the Ising model and $N_t=24$ with $t=3.75$ for the $XZ$ Heisenberg model. This Trotter formula is associated to specific rotations over the Pauli axes defined in the Hamiltonian exponents. Finally, the circuit is transpiled by means of Qiskit, which results in using two CZ gates per two-body interaction, setting the circuit depths and two-qubit gate counts described in the main text.

\subsection{Relative error for each simulation}\label{Sec: RelErr}

\begin{figure}
    \begin{overpic}[width=0.49\textwidth]{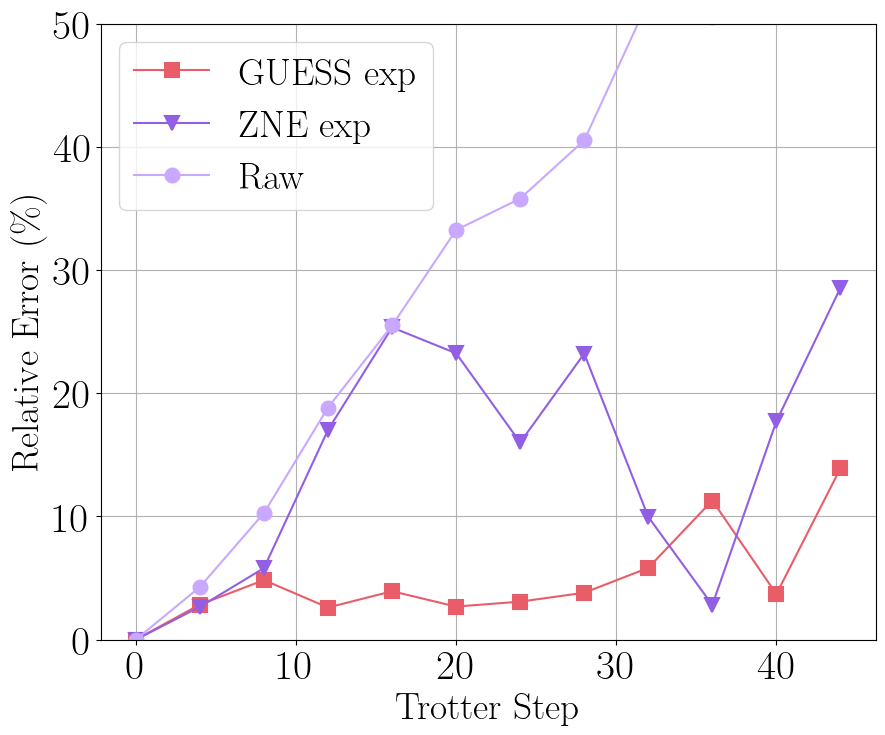}\put(0,80){(a)}
    \end{overpic}\hfill
    \begin{overpic}[width=0.49\textwidth]{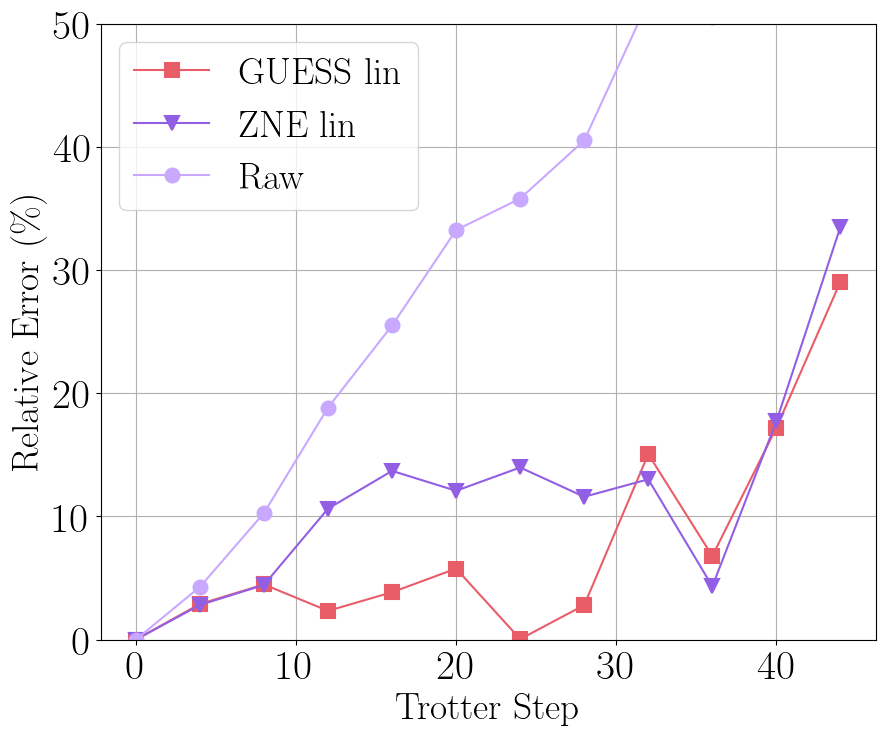}
    \end{overpic}
    \medskip
    \medskip
    \begin{overpic}[width=0.49\textwidth]{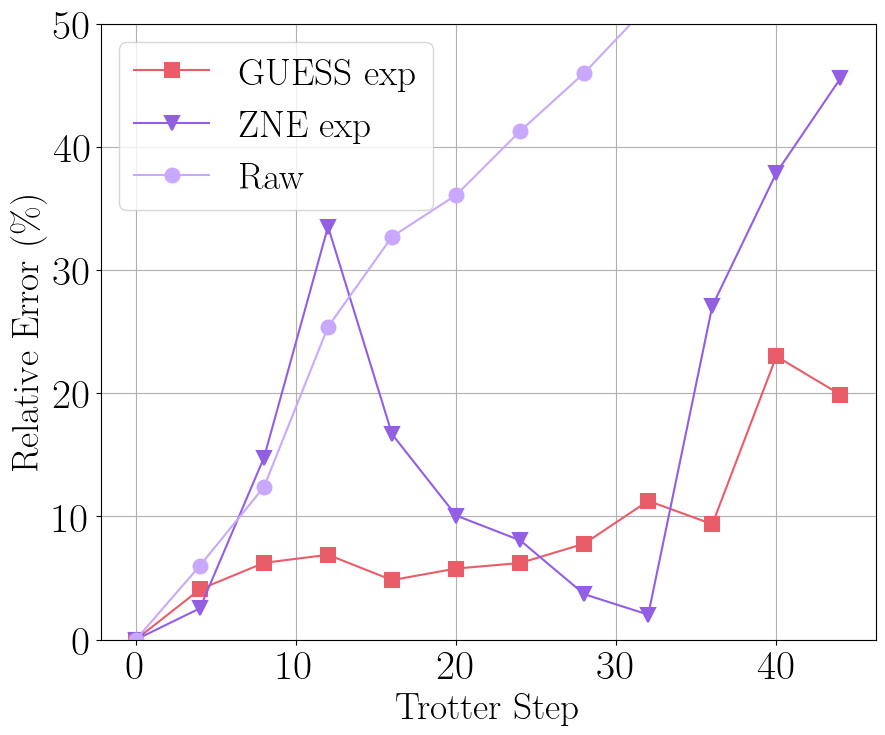}\put(0,80){(b)}
    \end{overpic}\hfill
    \begin{overpic}[width=0.49\textwidth]{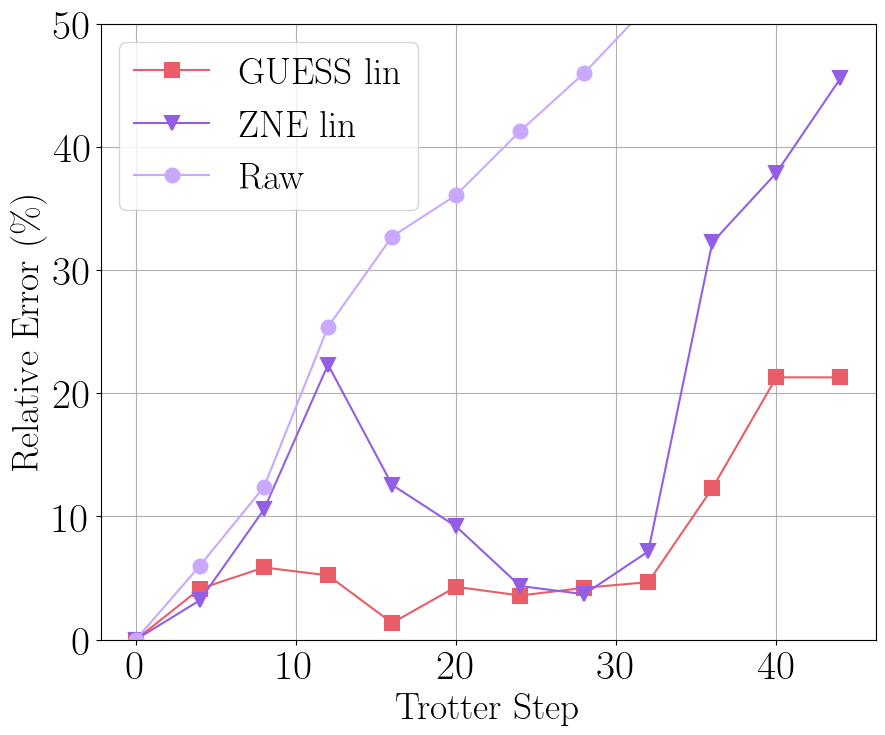}
    \end{overpic}
    \begin{overpic}[width=0.49\textwidth]{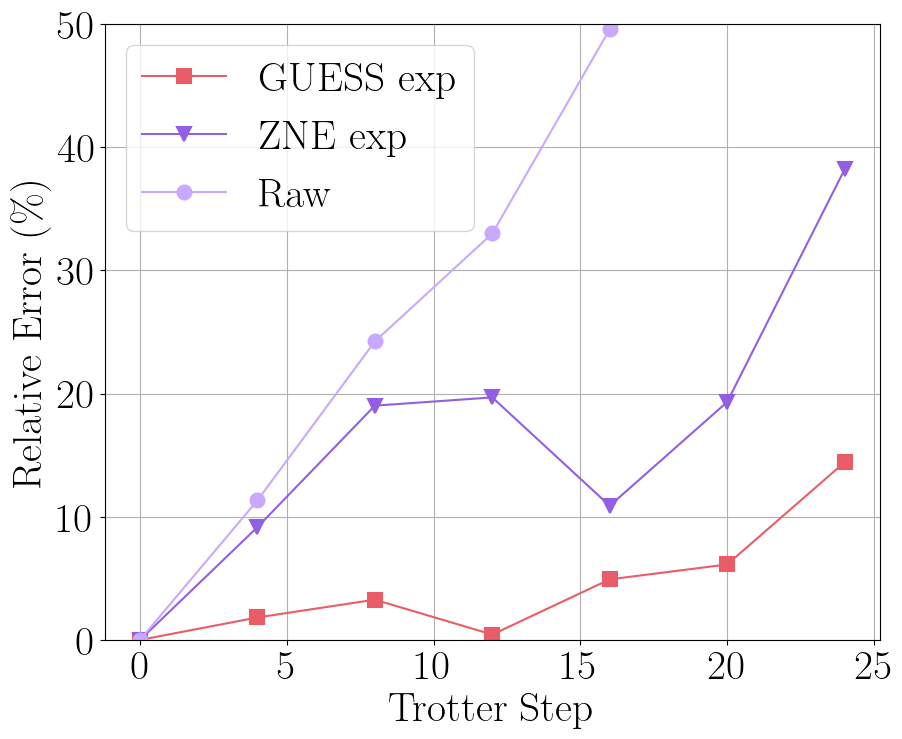}\put(0,80){ (c)}
    \end{overpic}\hfill
    \begin{overpic}[width=0.49\textwidth]{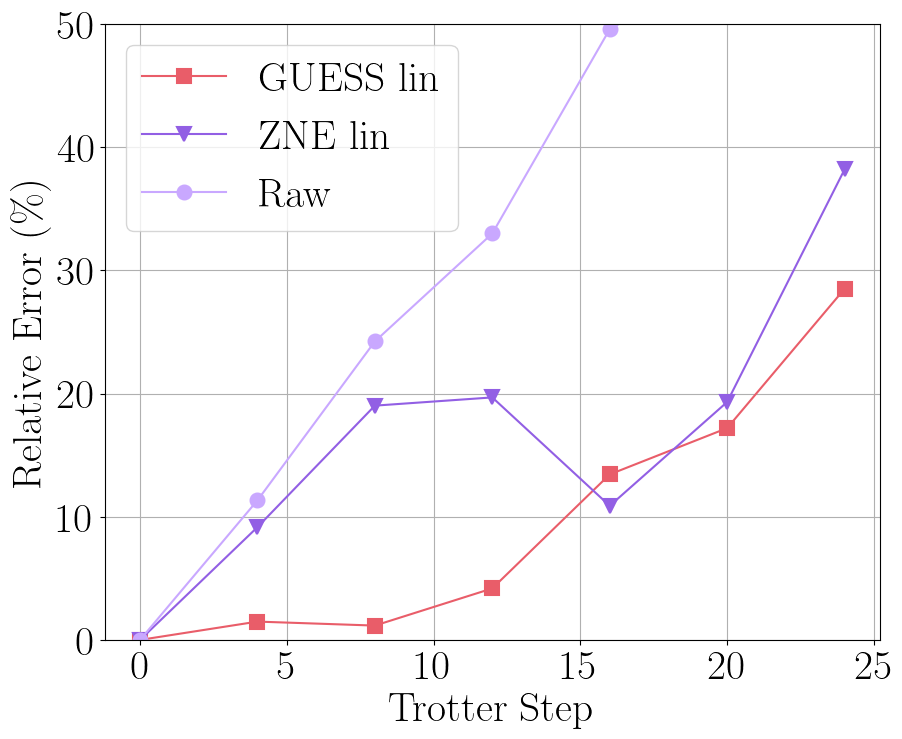}
    \end{overpic}
    \caption{Relative error of (a) the average magnetization and the (b) average nearest-neighbor correlators over 20 observables for the Ising model; and of (c) the average magnetization of the $XZ$ Heisenberg model over 20 observables.}
    \label{fig:placeholder}
\end{figure}

Here, we present the relative error for each of the models and simulations. We have computed the relative error for ZNE and GUESS as:
\begin{align}
    \text{Error}_{rel} (\%)= 100\cdot\frac{|\langle\bar{O}_{\text{MPS}}\rangle-\frac1N \sum_{i=0}^{N-1}\langle O^{\text{mit}}_i\rangle|}{\langle\bar{O}_{\text{MPS}}\rangle} 
\end{align}
The results for the Ising model for the averaged magnetization and the nearest-neighbor correlators are shown in Fig.~\ref{fig:placeholder}a) and b). In Fig.~\ref{fig:placeholder}c) the relative error is shown for the average magnetization of the $XZ$ Heisenberg model. Here, it is easy to see how GUESS gives more accurate values to the ideal ones and its consistency along the trotter steps, \textit{i.e.} the error increases almost constantly each trotter step when ZNE does not. 

\section{Diamond distance bounds for noise similarity after circuit perturbation} \label{Sec: Depol}

In the perturbation technique, we substitute certain gates in order to be able to learn the GUESS coefficients. For the sake of robustness, it is important to discuss that this perturbation does not significantly change how the noise propagates in the system. Here, we discuss this by means of the diamond norm distance~\cite{EtxezarretaMartinez2021,diamondnorm,watrous}. Specifically, we will bound the quantity $\lVert U_1\circ\mathcal{N}\circ U_1^\dagger - U_2\circ\mathcal{N}\circ U_2^\dagger\rVert_\diamond$, where $U_1$ and $U_2$ refer to the target and perturbed unitary evolutions in the circuit. The interpretation of the derived quantity is to compare the action of the propagated noise channel to the input state without taking into account the action of the unitaries to the state, \textit{i.e.} $U\circ\mathcal{N}\circ U^\dagger = \mathcal{N}_{U}\circ U\circ U^\dagger=\mathcal{N}_{U}$, where $\mathcal{N}_{U}$ is the propagated noise. From a physical perspective, this can be seen as the interaction picture. Taking this into account, we can bound the diamond norm as
\begin{align}
    \lVert U_1\circ\mathcal{N}\circ U_1^\dagger - U_2\circ\mathcal{N}\circ U_2^\dagger\rVert_\diamond =& \rVert\mathcal{N} - U_1^\dagger\circ U_2\circ\mathcal{N}\circ U_2^\dagger\circ U_1\rVert_\diamond
    \leq
    \lVert\mathcal{N}-\mathcal{I}\rVert_\diamond + \lVert\mathcal{I}-U_1^\dagger\circ U_2\circ\mathcal{N}\circ U_2^\dagger\circ U_1\rVert_\diamond
    \\
    =&\lVert\mathcal{N}-\mathcal{I}\rVert_\diamond + \lVert\mathcal{I}-\mathcal{N}\rVert_\diamond
    = 2\lVert\mathcal{N}-\mathcal{I}\rVert_\diamond,
\end{align}
where $\mathcal{I}$ refers to the identity channel. The first equality follows from the unitary invariance of the diamond norm while the inequality arises from the triangle inequality \cite{watrous,diamondnorm}, concretely: $\lVert\mathcal{N_\phi}-\mathcal{N_\psi}\rVert_\diamond = \lVert\mathcal{I}\circ\mathcal{N_\phi}-\mathcal{N_\psi}\circ\mathcal{I}\rVert_\diamond\leq\lVert\mathcal{I} - \mathcal{N_\psi}\rVert_\diamond + \lVert\mathcal{N_\phi}\circ\mathcal{I}\rVert_\diamond$. Considering now that the channel $\mathcal{N}$ is a Pauli channel whose diamond norm distance is well known, we can bound the quantity of interest as
\begin{align}
    \lVert U_1\circ\mathcal{N}\circ U_1^\dagger - U_2\circ\mathcal{N}\circ U_2^\dagger\rVert_\diamond
    \leq 2\lVert\mathcal{N}-\mathcal{I}\rVert_\diamond
    = 2\left(|1-(1-p)|+\sum_{P \in \mathcal{P}^{\otimes n}}p_P\right) = 4p,
\end{align}
where $p=\sum_{P \in \mathcal{P}^{\otimes n}}p_P$ is the total error probability of the Pauli channel in question. If the two-qubit gate error probability $p\sim10^{-3}$~\cite{UtilityIBM}, then the difference between the two propagated channels is bounded by a few times $10^{-4}$, which is small. Therefore, the target and perturbed propagated channels are close in terms of the diamond norm distance, and we can conclude that the perturbation technique used to derive the GUESS parameters does not have a significant impact on the noise structure of the circuit, which is later confirmed by the experiments with real hardware.

\section{Variance of extrapolation methods}  \label{Sec: GUESS variance}

\subsection{GUESS variance}
The GUESS method described along this work is composed by two different steps. Here, we will analyze how the variance is propagated to the mitigated observables. First, the $x_i$ coefficients are learned from an observable, $S$, that commutes with the Hamiltonian satisfying:

\begin{align}
    b_{s;i} = \sum  M_{s;i,j}\textrm{x}_j.
\end{align}
where $b_{s;i} \equiv \Tr(S_i \rho(0))$ is a known value (from the symmetry) and the sum is over the expectation values at different hardware noise rates. However, the expectation values in $M_{i,j}$ are obtained from a finite number of quantum measurements and therefore have an associated standard deviation, $\sigma_{i,j}$. We compute the coefficients $\textrm{x}_i$ using the minimum-norm pseudo-inverse solution as the system is undetermined and the uncertainty in the measurements propagates to the estimated coefficients. To quantify this propagation in the standard deviation we compute the Jacobian
\begin{align}
    J_{i, (j,k)} = \frac{\partial \textrm{x}_i}{\partial M_{s;j,k}},
\end{align}
which determines the sensitivity of each parameter to the uncertainty of each noisy expectation value. Then, the covariance of the coefficients is given by:
\begin{align}
    \textrm{Cov}(\vec{\textrm{x}})_{i,\ell} = \sum_{j,k} J_{i, (j,k)} \sigma^2_{j,k} J_{\ell, (j,k)},
\end{align}
where $\sigma^2_{j.k}$ are the diagonal terms of the covariance matrix of the noisy expectation values. Finally, the variances of the coefficients are formed by the diagonal entries of $\textrm{Cov}(\vec{\textrm{x}})_{i,\ell}$. 

Second, we use the coefficients to estimate the noiseless expectation value of the target observable. Again, the noisy expectation values have an associated uncertainty. To get the uncertainty after the whole mitigation we have to take into account the uncertainty in the learned coefficients. The model now is defined as:

\begin{align}
    b_i = \sum M_{i,j}\textrm{x}_j.
\end{align}
where $b_i = \Tr(O_i \rho_0)$ is the noiseless expectation value which we aim to mitigate. Its variance based on this extrapolation model is given by:
\begin{align} \label{Eq: varGUESS}
    \textrm{Var}(b_i)= \sum_j M_{i,j}^2\sigma_{x_j}^2  + \sum_j x_j^2 \sigma_{M_{i,j}}^2 + \sum_j \sigma_{x_j}^2 \sigma_{M_{i,j}}^2
\end{align}

where $\sigma_{x_j}^2$ and $\sigma_{M_{i,j}}^2$ are the variances of the coefficients and the expectation values, respectively. The first and the second terms account for the contribution of the variance of $\vec{\textrm{x}_j}$ and $M_{i,j}$ to the uncertainty of the noiseless expectation values. The last term in the expression is a high order term which accounts for the overlap of both uncertainties. For the exponential GUESS method, the parameters are learned from:

\begin{align}
    b_i = \exp\left(\sum M_{i,j}\textrm{x}_j\right)
\end{align}

The problem is linearized by taking the natural logarithm and solving the it as above. The propagation of the uncertainties to the coefficients is computed similarly. In the second step, we take the natural logarithm again and the variance is given by:

\begin{align}
    \textrm{Var}(\ln b_i)&=\sum_j M_{i,j}^2\sigma_{x_j}^2  + \sum_j x_j^2 \sigma_{M_{i,j}}^2 + \sum_j \sigma_{x_j}^2 \sigma_{M_{i,j}}^2
\end{align}

Taking the first order Taylor expansion of the exponential:

\begin{align}
    \textrm{Var}(b_i) = b_i^2 \textrm{Var}(\ln b_i)
\end{align}

Thus, we can see that the actual variance of the proposed QEM method does not depend on the uncertainty of the noise scaling parameters. It is related with how close the decay of the symmetries are to the target observables. As discussed in the text, this will be reasonable when using the same scaling method, if the noise in the system is similar both for the perturbed and the target circuits (maintaining the two-qubit depth) and if the impurity does not have a high impact in how the noise is propagated through the system (see Appendix~\ref{Sec: Depol}).

\subsection{ZNE variance} \label{Sec: ZNE variance}

ZNE can be performed using different extrapolation models, being the linear and exponential versions the most widely employed. The linear extrapolation propagates the uncertainty in the expectation value and its exact noise scale to the extrapolated value as:
\begin{align}
    \sigma^2_0= \sigma^2_{\text{res}} \left[\frac1n + \frac{\bar{x}^2}{S_{xx}}\right],
\end{align}
where $\sigma^2_{\text{res}}= \frac{1}{n-2}\sum_i \left( y_i - a - b x_i\right)^2$ is the variance of the residuals, $\bar{x}$ is the mean, and $S_{xx}= \sum_i \left(s_i-\bar{x}\right)^2$ is a measure of the spread of the noisy expectation values. In the case of the exponential model, the problem can also be linearized by means of logarithms, similar to what we discussed for the GUESS method. In the exponential space, the variance is given by:

\begin{align}
    \sigma^2_0 = y_0^2\sigma^2_{\text{res}} \left[\frac1n + \frac{\bar{x}^2}{S_{xx}}\right],
\end{align}
where, now, the variance of  the residuals is given by:
\begin{align}
    \sigma^2_{\text{res}} = \frac{1}{n-2} \sum_i \left( \ln y_i - \ln a - bx_i\right).
\end{align}

As expected, ZNE is sensitive to the uncertainty of the noise scaling technique due to the explicit dependence of the extrapolation model parameters to the actual noise gains considered. Since such scaling is a function of the actual ability of the noise amplification technique to provide accurate noise gains, the method experiences significant degradation in relative error and mitigation variance when the noise amplification technique is inaccurate.

\subsection{Overshoot - Non Physical results} \label{Sec: Non-physical}
As discussed before, imperfect noise scaling can result in ZNE methods to be pretty inaccurate, and cases in which the symmetry circuits are unable to replicate the decay of a target observable result in similar degradation for GUESS. The second scenario mainly occurs when noise drifts significantly from run to run~\cite{EtxezarretaMartinez2021} or when the system is too noisy to provide anything useful about the decay. A result to this effects is that the mitigation methods might provide estimated values with non-physical meaning, \textit{i.e.} their expectation value is greater than $1$ or lower than $-1$. In this cases, where the result is obviously incorrect, a mitigation method can be changed to another one (e.g. from an exponential GUESS to a linear GUESS) or to the raw value. In Table~\ref{Table: Non-phys} we compare the number of non physical noiseless expectation values for all the scenarios we have analyzed in this work for real hardware implementation. For our results, we have taken these non-physical values into account. Thus, when we refer to an exponential version, the method degrades to a linear if overshoot occurs, conversely for the linear models, and to the raw if again the value is non-physical. This type of hierarchical mitigation is typical in the literature~\cite{UtilityIBM} whenever one is confident that the mitigated values are wrong.

We can observe that whenever a more aggressive extrapolation technique is considered, \textit{i.e.} the exponential one, the proportion of non-physical values reported by it increases consistently both for GUESS and for ZNE. Importantly, note how the exponential ZNE estimates non-physical values in most of the cases. The main reason is the fact that ZNE relies on the certainty of the noise scaling factors and a small uncertainty about their value has a high impact in the extrapolated value and its variance (see appendix~\ref{Sec: ZNE variance}). Partial gate folding does not scale the noise accurately and other methods, as PEA, requires a good characterization of the noise~\cite{UtilityIBM}. However, even this methods present a small uncertainty which increases over time of computation due to the noise drift. The variance of the mitigated value using GUESS method comes from the uncertainty in the expectation values of the symmetry, which propagates to the learned coefficients, and then from the uncertainty of the noisy expectation values of the target observable. But is is not affected by uncertainties in the noise scaling. Instead, its performance is reduced if the noise scaling changes between the run of the Hamiltonian with and without the impurity, thus, we expect better results for an improvement in the noise scaling technique.

\section{Symmetry outliers detection} \label{Sec: Postpro}

\begin{figure}
    \begin{overpic}[width=0.49\textwidth]{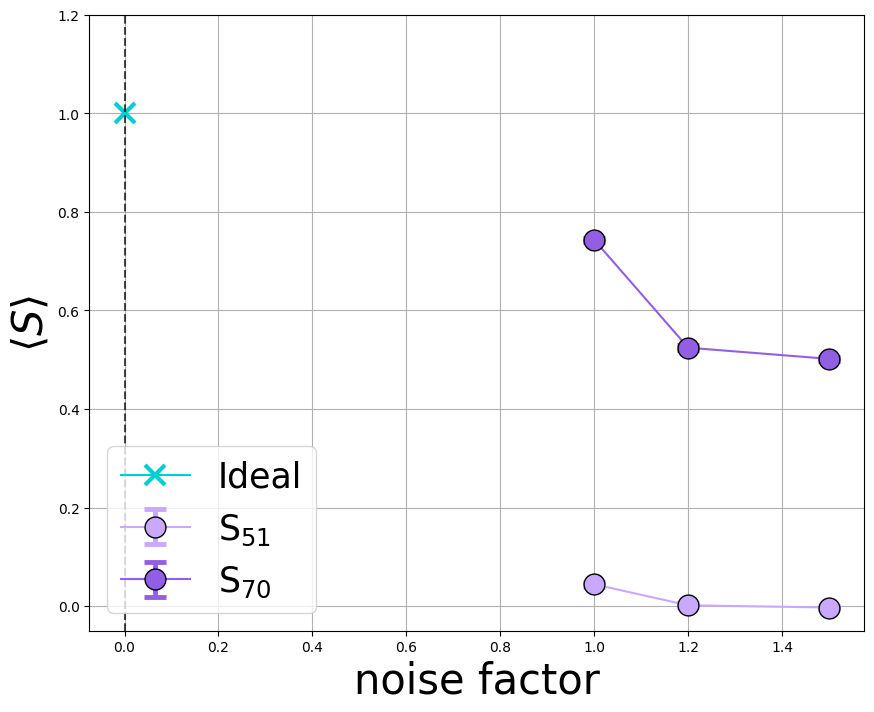}\put(0,80){(a)}
    \end{overpic}\hfill
    \begin{overpic}[width=0.49\textwidth]{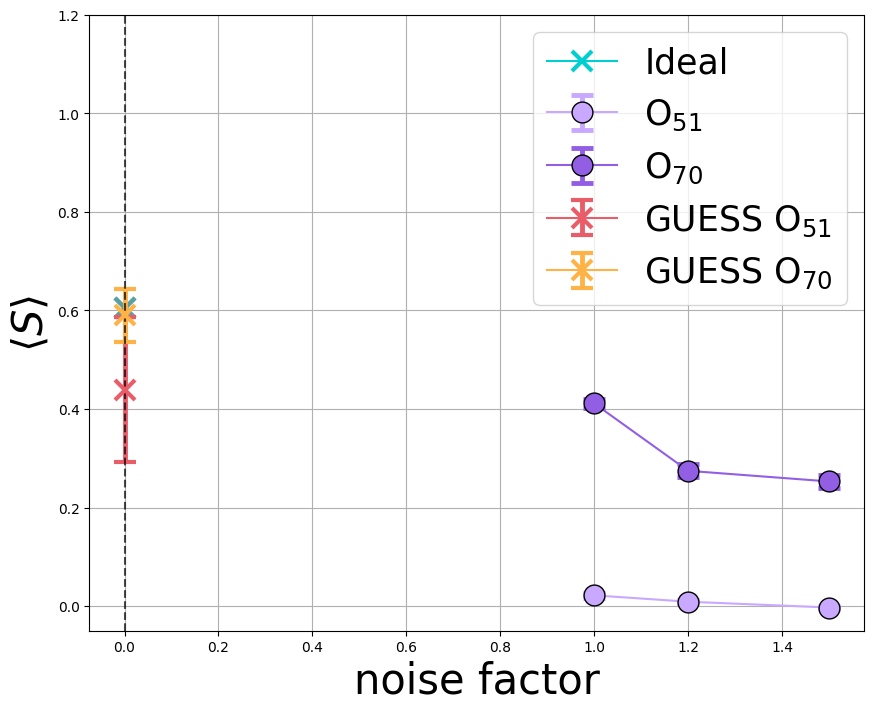}
    \end{overpic}
    \medskip
    \caption{We show two representative examples obtained from the simulation of the TFIM to get the average magnetization at trotter step $24$. In the left figure we represent the symmetries $S_{51}$ and $S_70$, corresponding to the qubits $51$ and $70$, from which we learn the GUESS coefficients to mitigate the target observables $O_{51}$ and $O_{70}$, respectively. From the simulation of the Hamiltonians with the impurity, it is evident that the qubit $51$ experiences significantly stronger decoherence compared to qubit $70$. In the right panel, we can observe the mitigated expectation value using the GUESS method. As for the perturbed Hamiltonian, the qubit $70$ suffered more decoherence resulting in a less accurate value with larger variance compared to the value recovered from qubit $51$.}
    \label{fig:goodvsbad}
\end{figure}

As discussed in the main text, GUESS requires extra circuit runs using the perturbed Hamiltonian (see Eq.~\eqref{Eq: ImpHamil}) which are then used to learn the extrapolation coefficients. This increases the overhead when compared to standard ZNE (twice circuit runs per observable), but it provides useful information on the behavior of noise in the system. In fact, we can use the information of the measured symmetry to perform more tasks than just learning the extrapolation coefficients. For instance, it introduces the possibility of determining when a qubit is functioning defectively, to the extent that the measured expectation values become too unreliable even for error mitigation. In this work, we have utilized this fact by measuring a total of $40$ observables across both the perturbed and target Hamiltonians. Note that by only studying the noisy values of target circuit, we cannot decide if a circuit run is too noisy since the ideal behavior is unknown. However, for the case of the perturbed Hamiltonian we exactly know how it should behave. Thus, it brings the possibility of characterizing the quality of a qubit because we can know if it is not working as it should by analyzing the data taken from the symmetry. More specifically, when the noise is completely uniform across a device, the expectation values of the symmetries should be exactly the same for each qubit in each trotter step. In contrast, this is not what is observed experimentally, as the noise is extremely heterogeneous and drifts over time~\cite{EtxezarretaMartinez2021}, as shown in Fig.~\ref{fig:goodvsbad}. Therefore, we have used the extra information gathered from the symmetries in a two-fold way to post-process unreliable expectation values:
\begin{itemize}
    \item The uncertainty of the machine measured expectation values depends on the number of circuit shots used to estimate it and on the variance introduced by the Pauli twirl process since this is done by randomizing circuit runs in practice~\cite{Twirl1,Twirl2}. The symmetric observable allows us to analyze its results look for the outliers, \textit{i.e.} cases for which the uncertainty of the raw measurements is too high. We can study their distribution to identify the qubits that have an outlier standard deviation compared to what is expected. Specifically, we filtered out $10$ expectation values considered outliers based on their standard deviation, starting from the shortest circuits and continuing until reaching the maximum. We used the interquartile range (IQR) for the post-selection as: $Q3 + 1.5$IQR, where IQR$=Q_3 - Q_1$ and $Q_i$ is the i$^{th}$ quartile.
    \item Extrapolation techniques, and QEM techniques in general, have a strong dependence on how noisy is the raw noisy expectation value of certain observable~\cite{Cai_2023}. Thus, we can also use the symmetries to determine a measured expectation value is too noisy as the ideal value is completely known. The reported results in the main text present the average of the $20$ best observables in each case based on this fact, \textit{i.e.} we discard the other $10$ with worst bias in the symmetry case.
\end{itemize}

A pseudo-code of this post-selection process is given in Algorithm~\ref{Pseudo: Outliers}. Our primary objective in this work is to validate the efficacy of the proposed QEM method by identifying and excluding statistical outliers. However, in applications where all observables are deemed physically significant for the final outcome, we can make use of a post-processing technique as an iterative refinement process to re-evaluate the outlier instances until the non-outlier criteria is satisfied and we can accept them. Thus, we consider that GUESS allows for extensive validation of the quality of the measured samples, allowing for extensive refinement of QPU mitigated quantum observables. Moreover, more refined statistical post-selection techniques may be applied to assess the quality of the observables based on measured symmetries.

\begin{figure*}[ht]
    \begin{overpic}[width=0.49\textwidth]{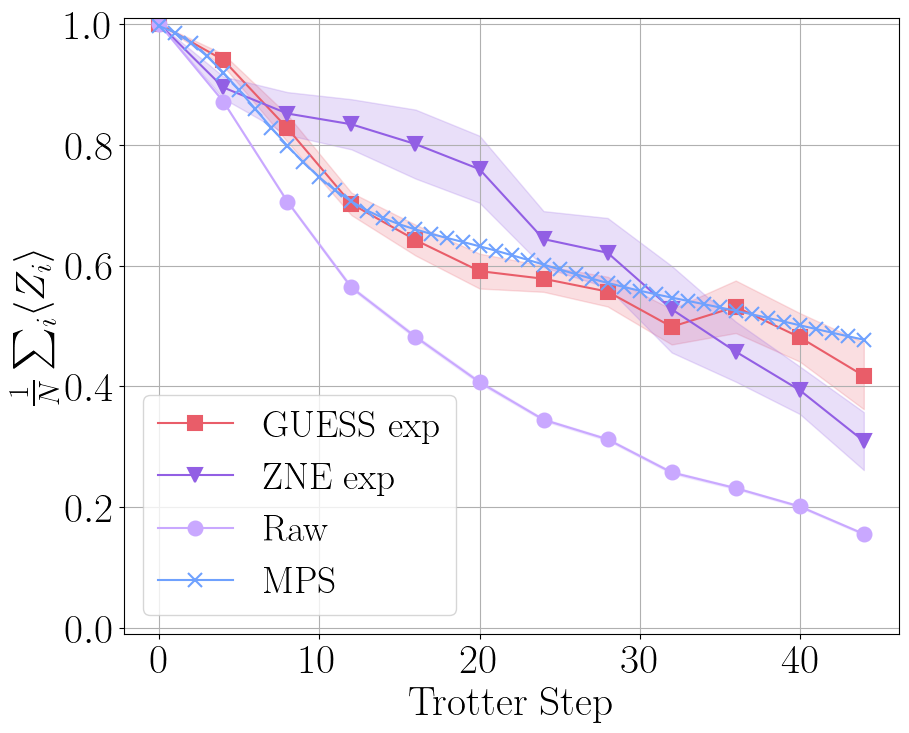}\put(2,80){\small (a)}
    \end{overpic}\hfill
    \begin{overpic}[width=0.49\textwidth]{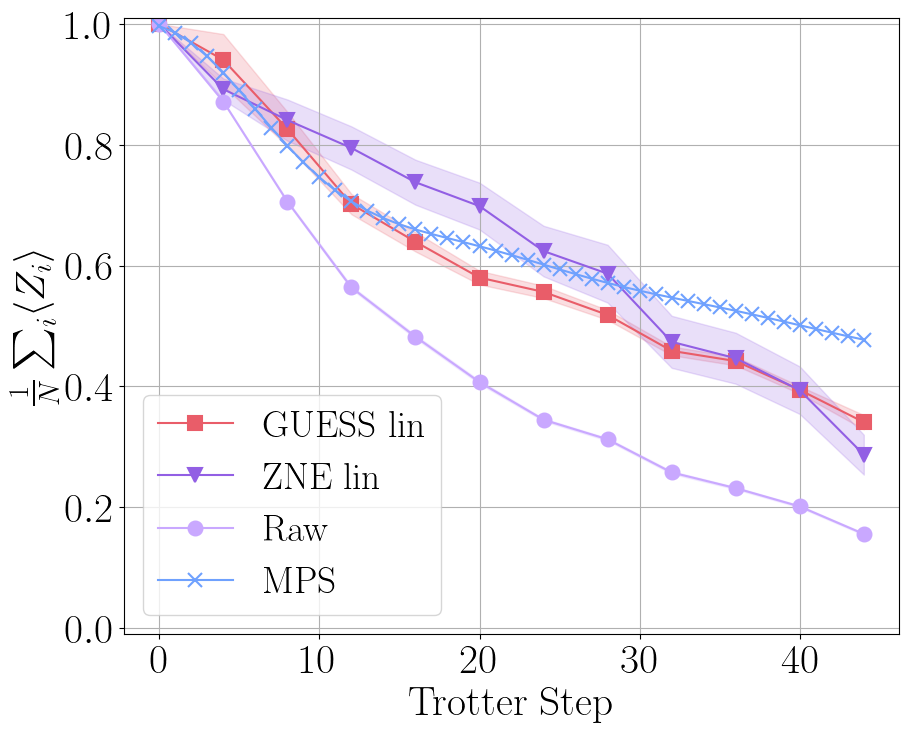}
    \end{overpic}
    \medskip
    \medskip
    \begin{overpic}[width=0.49\textwidth]{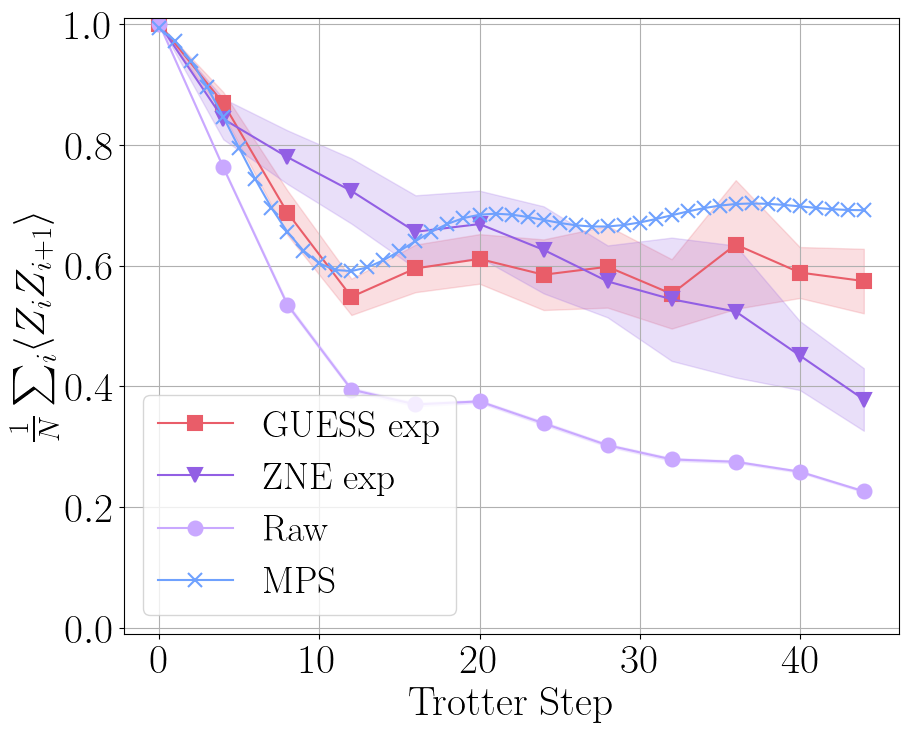}\put(2,80){\small (b)}
    \end{overpic}\hfill
    \begin{overpic}[width=0.49\textwidth]{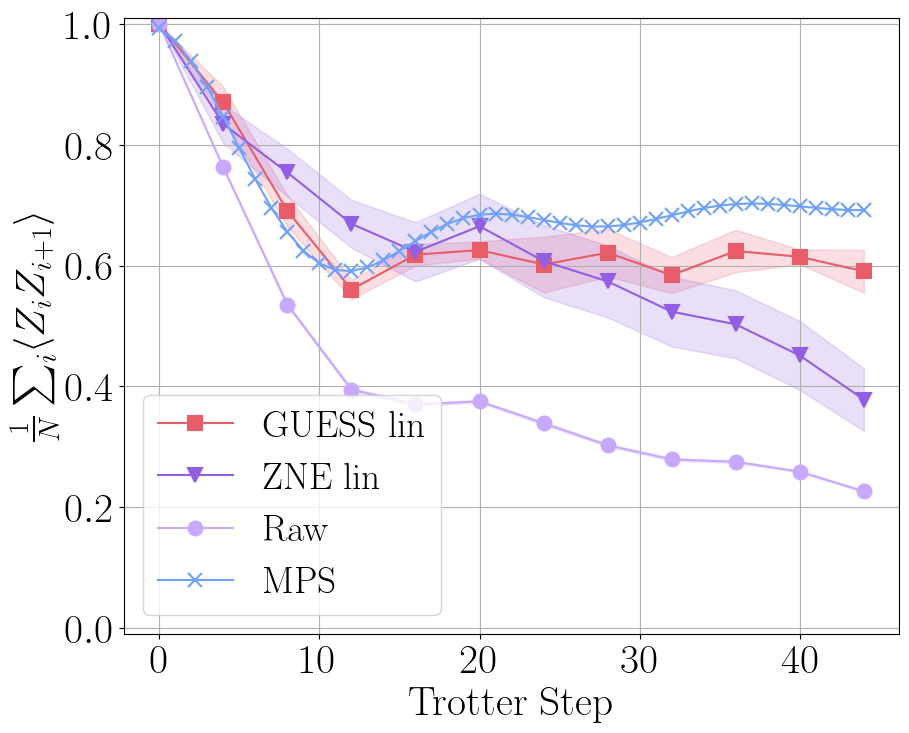}
    \end{overpic}
    \caption{Simulation of the considered 1D transverse field Ising model with open boundaries. We run a simulation up to $44$ trotter steps measuring each $4$ steps. The circuit has a maximum depth of $176$ and a total number of $8712$ CZ gates for the baseline case $G=1$. We show the result for all measured $40$ observables based on our statistical postprocessing process (see Appendix~\ref{Sec: Postpro}) for the exponential and the linear models, left and right columns, respectively. In panel (a) we show the dynamics of the average magnetization (weight-$1$ observables) and in panel (b) of the average nearest-neighbor correlators (weight-$2$ observables). The blue points (cross) refer to the ideal values, obtained from the MPS simulation; the red (square) points to the extrapolated values using GUESS; and the dark (triangle) and light (circle) to ZNE and the raw machine measurements, respectively.}
    \label{fig: Ising-wt1-all}
\end{figure*}

\begin{figure*}[ht]
\includegraphics[width=0.5\textwidth]{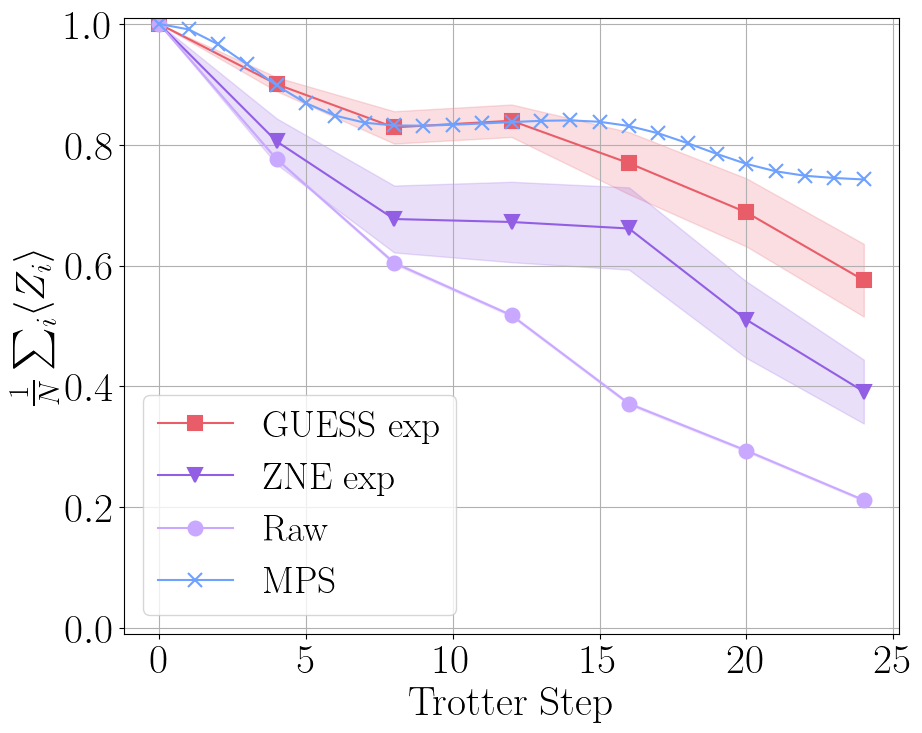}\includegraphics[width=0.5\textwidth]{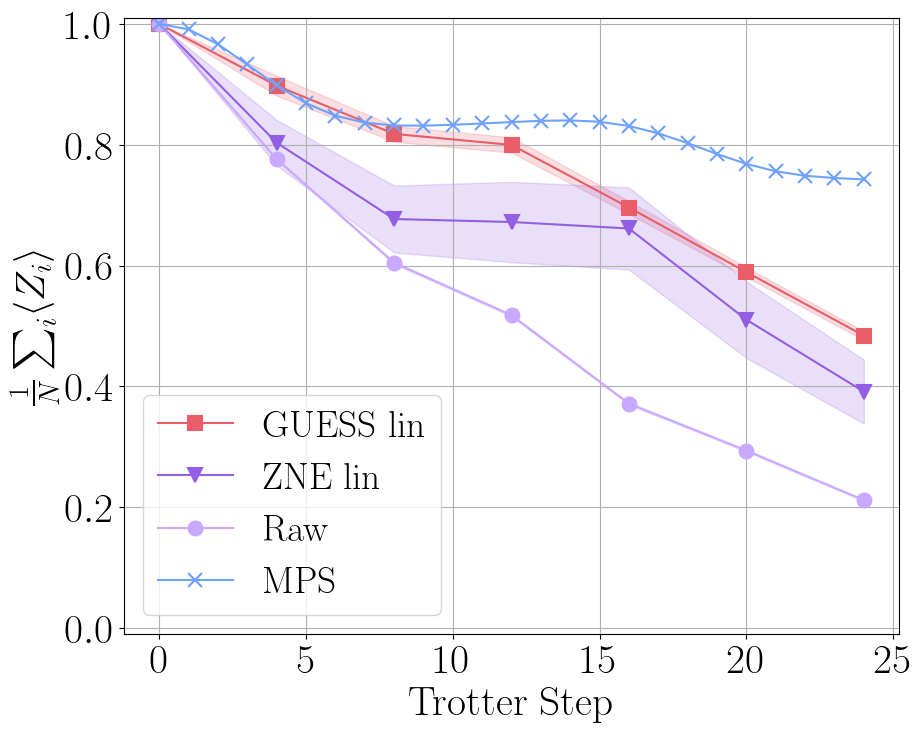}
    \caption{Simulation of the considered 1D transverse field $XZ$ Heisenberg model with open boundaries. We run a simulation up to $24$ trotter steps measuring each 4 steps. The circuit has a maximum depth of $192$ and a total number of $9504$ CZ gates for the baseline case $G=1$. We show the result for all measured $40$ observables based on our statistical postprocessing process (see Appendix~\ref{Sec: Postpro}) for the exponential and the linear models, left and right columns, respectively. We show the averaged magnetization, where the blue points (cross) refer to the ideal values, obtained from the MPS simulation; the red (square) points to the extrapolated values using GUESS; and the dark (triangle) and light (circle) to ZNE and the raw machine measurements, respectively.}
    \label{fig: Heis-all}
\end{figure*}

Anyway, in Figures~\ref{fig: Ising-wt1-all} and \ref{fig: Heis-all}, we show the results without this post selection process (taking the average over the $40$ observables selected based on the best readout) for the sake of completeness. It can be seen that GUESS is still accurate at mitigating noisy quantum observables and superior to standard ZNE. Note, however, how the additional less trustworthy observables make the relative error and the variance to increase compared to the results in the main text.

\begin{algorithm} 
\caption{Post-Selection of Quantum Observables \label{Pseudo: Outliers}}
\begin{algorithmic}[1]

\Require Set of Symmetries $\{S\}$, Experimental Data (Expectation Values $M_{\mathrm S}$, Standard Deviations $\sigma_{i,j}$)

\Procedure{PostSelection}{$\{S_i\}, M_{\mathrm S}, \sigma_{i,j}$}
    
    \State \textbf{Step 1: Outlier Removal based on Noise ($\sigma_{i,j}$)}
    \For{each Trotter step $t$}
        \State Calculate statistics ($Q_1, Q_3, IQR$) of $\sigma_{i,0}$ for all valid observables
        \State Define noise threshold: $Limit = Q_3 + k \times IQR$
        \For{each observable $S_k \in\{S\}$}
            \If{$\sigma_{k,0} > Limit$}
                \State Mark $k$ as \textbf{Bad} qubit
            \EndIf
        \EndFor
    \EndFor
    
    \State \textbf{Step 2: Data Cleaning}
    \For{each symmetry $S_k \in \{S\}$}
        \If{$k$ is marked \textbf{Bad}}
            \State Discard $M_{\mathrm S,k}$ and $\sigma_{k,j}$ for all steps (set to NaN)
        \EndIf
    \EndFor

    \State \textbf{Step 3: Selection based on Expectation Values ($M_{\mathrm S}$)}
    \State Filter remaining valid symmetries $S_\text{good}$
    \State Select the best $20$ qubits $k$ in $S_{best}$ with highest expectation value
    
    \State \Return Best qubits associated to $S_{best}$
\EndProcedure

\end{algorithmic}
\end{algorithm}

\end{document}